\documentclass[prc,twocolumn,showpacs,floatfix,nofootinbib,preprintnumbers,superscriptaddress,amsmath,amssymb]{revtex4}

\usepackage{CJKutf8}
\usepackage{graphicx}
\usepackage{epsfig}
\usepackage{amsmath,amssymb}
\usepackage{bm}
\usepackage{color}
\usepackage{float}
\usepackage{dcolumn}
\usepackage{multirow}
\usepackage{mathrsfs}
\usepackage[arrowdel]{physics}
\usepackage[OT1]{fontenc}
\usepackage{outline}
\usepackage{tikz,tikz-3dplot}
\usetikzlibrary{positioning}

\DeclareMathAlphabet{\mathcal}{OMS}{cmsy}{m}{n}

\renewcommand{\vec}[1]{\mbox{\boldmath $#1$}}

\newcommand{\HFODD}{\textsc{hfodd}}
\newcommand{\HFBTHO}{\textsc{hfbtho}}
\newcommand{\UNEDFone}{\textsc{unedf1}}

\begin{document}
\begin{CJK*}{UTF8}{gbsn}

\title{An implementation of nuclear time-dependent density-functional theory
and its application to the nuclear isovector electric dipole resonance}

\author{Yue Shi (石跃)\thanks{corresponding author}}
\email[corresponding author: ]{yueshi@hit.edu.cn}
\affiliation{Department of Physics, Harbin Institute of Technology, Harbin 150001, People's Republic of China}

\author{Nobuo Hinohara}
\affiliation{Center for Computational Sciences, University of Tsukuba, Tsukuba 305-8577, Japan}
\affiliation{Faculty of Pure and Applied Sciences, University of Tsukuba, Tsukuba 305-8571, Japan}

\author{Bastian Schuetrumpf}
\affiliation{GSI Helmholzzentrum f\"ur Schwerionenforschung, Planckstra\ss{}e 1, 64291 Darmstadt, Germany}

\begin{abstract}
\begin{description}
\item[Background] Time-dependent density-functional theory (TDDFT)
continues to be useful in describing a multitude of low-energy static and
dynamic properties. In particular, with recent advances of computing
capabilities, large-scale TDDFT simulations are possible for fission dynamics
as well as isovector dipole (IVD) resonances.

\item[Purpose] Following a previous paper [Y. Shi, Phys. Rev. C {\bf 98},
014329(2018)], we first present a time-dependent extension of the
density-functional theory to allow for dynamic calculations based on the
obtained static Hartree-Fock + Bardeen-Cooper-Schrieffer (BCS) results. Second,
we apply the TDDFT + BCS method to a systematic description of the IVD
resonances in the Zr, Mo, and Ru isotopes.

\item[Methods] To benchmark the TDDFT code, we compute the strengths of IVD
resonances for light nuclei using two complementary methods: TDDFT and FAM-QRPA
methods. For the TDDFT results, additional benchmark calculations have been
performed using the well-tested code Sky3D. In these three calculations, the
important ingredients which have major influence on the results, such as
time-odd potentials, boundary conditions, smoothing procedures, spurious peaks
etc., have been carefully examined.

\item[Results] The current TDDFT and the Sky3D codes yield almost identical
response functions once both codes use the same time-odd mean fields and
absorbing boundary conditions. The strengths of the IVD resonances calculated
using the TDDFT and FAM-QRPA methods agree reasonably well with the same
position of the giant dipole resonance.
Upon seeing a reasonable accuracy offered by the implemented code, we perform
systematic TDDFT + BCS calculations for spherical Zr and Mo isotopes near
$N=50$, where experimental data exist. For neutron-rich Zr, Mo, and Ru
isotopes where shape evolution exists we predict the photoabsorption cross
sections based on oblate and triaxial minima.

\item[Conclusions] The TDDFT + BCS code provides reasonable description for IVD
resonances. Applying it to the spherical Zr and Mo nuclei, a reasonable
agreement with experimental data has been achieved. For neutron-rich Zr
isotopes, the photoabsorption cross section based on the two coexisting minima
reflects the feature of the deformation of the minima. This suggests the
possiblity of obtaining additional information about the ground-state
deformation by comparing the GDR data with the TDDFT + BCS results.

\end{description}
\end{abstract}


\maketitle
\end{CJK*}

\section{Introduction}
\label{sec1}

Since its first numerical realizations in the late
70s~\cite{engel75,Bon76a,Cus76a}, the time-dependent density-functional theory
(TDDFT) continues to be useful in describing a variety of low-energy nuclear
static and dynamic properties, ranging from the linear response of nuclear
density, to the large-amplitude motion of heavy
nuclei~\cite{nege82,naka16,Umar15,Simenel12,burr19}.
The modern developments~\cite{naka05,Maruhn2005,Umar06a} allow for the
inclusion of the full original Skyrme energy density functional (EDF) in the 
framework of nuclear density functional theory (DFT). Hence,
the same EDFs obtained from the knowledge of the
static properties of nuclei can be applied in dynamic simulations without any
further approximation.

With advances of computing capabilities, nowadays one can perform TDDFT
simulations that were not possible even twenty years ago. For example, the
linear-response properties of medium or heavy nuclei, fission dynamics of
actinides, as well as nuclear reaction involving medium-heavy nuclei are within
the reach of calculations with single-node computers.

However, computational costs to include the nuclear pairing correlation in the
TDDFT dynamics is still very demanding. Several independent approaches with
different treatment of the pairing have been developed. For instance, The BCS
pairing~\cite{scamps18} and Superfluid Local Density
Approximation~\cite{stetcu11,bulgac16,magi17} are employed in for the
time-dependent Hartree-Fock-Bogoliubov (TDHFB) simulation in the
three-dimensional (3D) Cartesian coordinate space.  In the linearized limit,
the coordinate-basis canonical TDHFB method has been systematically applied for
the low-lying diople mode~\cite{ebat10,ebata14}. The harmonic-oscillator (HO)
and Lagrange-mesh based TDHFB calculations with finite-range Gogny
force~\cite{hash12,hash13} have become available, taking advantage of the fact
that the oscillation extends only in a relatively small region. The latter code
is also applicable to the collision dynamics~\cite{hash16,hash17}. Recently,
finite-amplitude method for quasiparticle random-phase approximation (FAM-QRPA)
calculations in the 3D Cartesian coordinate space have also
emerged~\cite{wash17}.

The goal of our project is to describe low-energy large-amplitude motion such
as nuclear fusion and fission using TDDFT. To this end, the 3D Cartesian
coordinate space calculation with proper treatment of the dynamical nuclear
pairing correlation is indispensable. Some of the earlier
works~\cite{stetcu11,bulgac16,magi17} have developed these features. As a first
step towards this goal, in the present work, we show an extension of an earlier
developed Skyrme-HFB code~\cite{shi18} on time-dependent capabilities with
fixed-occupation probabilities~\cite{maru14,Schuetrumpf2018}. The code is
represented in the 3D Cartesian coordinate space, using a light-weighted
finite-difference method for derivative operators. The code features an
interface with the HFODD code~\cite{doba97a,doba09,schunck17}, which is a
Skyrme-Hartree-Fock-Bogoliubov (HFB) code in a 3D HO basis. Such a flexible
code is desired to provide a reasonable alternative for future development.

As the first application, the current work provides systematic calculations for
the isovector (IV) electric dipole ($E1$) vibration motion for stable and
neutron-rich Zr, Mo, and Ru isotopes. Although there exist a few systematic
calculations for IV and isoscalar vibrational properties for nuclei across the
nuclear chart~\cite{inak11,scamps13,scamps14,ebata14}, we find that a detailed
analysis of the shape evolutions and shape coexistence in the same nucleus,
reflected by the different structures of the GDR cross sections, is
particularly useful~\cite{kvasil09}.

In Sec.~\ref{model} we present a description about the main features of the
current TDDFT + BCS framework. Section~\ref{results} contains two parts: first, a set
of careful benchmark calculations, with the current TDDFT, Sky3D codes, and the
FAM-QRPA calculations, have been presented. Second, systematic calculations
have been performed for the photoabsorption cross section of the isovector
dipole (IVD) vibration in the spherical and deformed Zr, Mo, and Ru nuclei. A
summary is contained in Sec.~\ref{summary}.

\section{The model}
\label{model}

This section describes in detail the procedure for the time development in
connection with the previous static calculation~\cite{shi18}. Then, we briefly
describe the Sky3D~\cite{maru14,Schuetrumpf2018} calculation, with which the
current code is benchmarked. For the applications in the linearized limit of
the current TDDFT calculation, we provide formulae for describing the relevant
properties associated with the $E1$ vibrational mode.

\subsection{The static calculations}

Before the single-particle wave functions are propagated in time, one has to
obtain the static solution of the Hartree-Fock (HF) problem. In this stage of
the calculation, the time-odd components of the densities and mean-fields
vanish for even-even nuclei. The form of the Hamiltonian, the way how the
operators of the Hamiltonian are constructed, and how the integrations are
performed have been explained in Ref.~\cite{shi18}.

\subsubsection{The grid points arrangement}

The grid points in the present implementation are moved away from the origin of
the simulating box and differs from those of Ref.~\cite{shi18}. Specifically,
in the example of one dimension, instead of using a set of coordinates at 
\begin{equation}
[-nx_{\rm max}, ...,0,1,...+nx_{\rm max}]\times dx,
\end{equation}
the current code represents the problem on grid points at the coordinates 
\begin{equation}
	\label{newgrid}
[-nx_{\rm max}+0.5, ...,-0.5,0.5,...+nx_{\rm max}-0.5]\times dx,
\end{equation}
where $nx_{\rm max}$ is an integer number numerating the points at the edge of
the simulating box. The $dx$ denotes the grid spacing. Note, that the latter
choice has an even number of grid points, whereas the former one has an odd
number of grid points. This choice is guided by the fact that the inclusion of
the grid point at the origin of the box results in numerical
problems~\cite{maru14}. Using the grid shown in Eq.~(\ref{newgrid}), the
integration can be carried out by summation {\it on} the grid, without the
interpolation as presented in Ref.~\cite{shi18}.

\subsubsection{The Bardeen-Cooper-Schrieffer (BCS) pairing}

To demonstrate the influence of the pairing interaction on the properties of
the IVD resonances, we include a simple BCS
pairing~\cite{bardeen57,ring80,bender00}. For the BCS method, we attach each
single-particle wave function a real number, $v_i$, whose square gives the
occupation probability of the $i$th orbit.

After each HF iteration, the occupation amplitude $v_i$ is determined, in the
current work, by the following BCS equations
\begin{equation}
\label{occupation}
v_{i,q}^2=\frac{1}{2}\left[1-\frac{\epsilon_{i,q}-\lambda_{q}}{\sqrt{(\epsilon_{i,q}-\lambda_q)^2+\Delta_{i,q}^2}}\right],
\end{equation}
where $\epsilon_{i,q}$'s are the HF single-particle energies; $\lambda_q$ is the
Fermi energy for given nucleonic type, which is adjusted so that $2\sum_i v_{i,q}^2$
gives the correct nucleon number. In Eq.~(\ref{occupation}), the
state-dependent single-particle pairing gaps, $\Delta_{i,q}$'s, are given by
\begin{equation}
\Delta_{i,q} = \sum_{\sigma}\int d \bm{r} \Delta_q(\bm{r}) \psi_{i,q}^*(\bm{r},\sigma)\psi_{i,q}(\bm{r},\sigma),
\end{equation}
where
\begin{align}
\label{densities_bcs}
\Delta_q(\bm{r}) &= -\frac{1}{2}V_q\int d \bm{r}\left[1-\frac{\rho(\bm{r})}{\rho_{\rm pair}}\right] \tilde{\rho}_q(\bm{r}), \\
\rho_q(\bm{r}) &= \sum_{i,\sigma}  v_{i,q}^2\, \psi_{i,q}^*(\bm{r},\sigma)\psi_{i,q}(\bm{r},\sigma), \\
\tilde{\rho}_q(\bm{r}) &= \sum_{i,\sigma}  v_{i,q}\, \sqrt{1-v_{i,q}^2}\, \psi_{i,q}^*(\bm{r},\sigma)\psi_{i,q}(\bm{r},\sigma).
\end{align}
with $q=n,p$ denoting the neutron and proton, respectively. The quantities
without subscripts denote the summed contributions from neutrons and
protons, for example, $\rho=\rho_n+\rho_p$. We choose $\rho_{\rm
pair}=0.32$\,fm$^{-3}$ in this work.

When applied to the drip-line nuclei, using the BCS pairing tends to scatter
the particles to the positive-energy levels which are non-local, resulting in
the unphysical nucleon gas surrounding the nucleus. This problem can be cured
by replacing the BCS theory with the HFB theory~\cite{doba84}. In the current
work, we limit the TDDFT+BCS calculations to the nuclei far from drip line.
This indicates that the Fermi surfaces are far from the positive-energy level.
Hence, these orbits are less occupied compared to those bound orbits. Among the
nuclei studied in the current work, the most neutron-rich one is $^{34}$Mg,
where the most significant occupation probability of the positive-energy level
is in the order of 10$^{-5}$, which is three orders of magnitude smaller than
the probability of the least occupied bound state. It has been checked that the
neutron densities for paired $^{34}$Mg would decrease exponentially with the
increase of the distance from the center of the nucleus. Hence, this small
occupation would not lead to the nucleon gas problem in the neutron density
distribution. For the neutron-rich Zr, Mo, and Ru isotopes, the largest
occupation of the scattering state is also in the order of 10$^{-5}$, for the
paired calculation shown in Sec.~\ref{zr_mo_ru}.

\subsection{The nuclear mean fields including the time-odd parts}

In the earlier static code presented in Ref.~\cite{shi18}, it has been
explained that the time-odd densities and mean fields vanish due to the
time-reversal symmetry. When time propagation is discussed, the time-odd
densities and mean fields appear~\cite{engel75}. Due to computing limitations,
historically, the earlier TDHF calculations contained a few serious
approximations such as the schematic treatment of the spin-orbit and pairing
interactions. Modern TDHF calculations~\cite{naka05,Maruhn2005,Umar06a} include
the full Skyrme interactions. Recent studies discuss the influence of the
tensor interactions when applied to the description of GDR~\cite{frac12} and
nuclear collisions~\cite{guo18}.

The current paper adopts the frequently-used Skyrme EDF which contains, in
addition to the time-even densities, time-odd densities $\bm{s}$ and $\bm{j}$. The
tensor interaction is not considered in this work. See Eq. (A.19) of
Ref.~\cite{engel75} for a detailed form of the Skyrme energy density
$\mathcal{H}(\bm{r})$.

After variation of the total energy, $E=\int\mathcal{H}(\bm{r})d \bm{r}$, with
respect to the density matrix, the resulting Skyrme mean fields also contain
terms of the above-mentioned time-odd densities. Modern nuclear DFT allows for
a free parametrization of coupling constants in front of each term in the
Skyrme mean field. See Eq. (2.6) of Ref.~\cite{doba95} for details. Assuming
local gauge invariance of the energy density, one requires the terms
contributing to the mean fields to be grouped in pairs~\cite{doba95},
specifically, ($\rho\tau-\bm{j}^2$) and ($\rho\bm{\nabla}\cdot
\bm{J}+\bm{s}\cdot{\bm{\nabla}\times\bm{j}}$).

In the current implementation of the TDDFT code, the single-particle Hamiltonian reads
\begin{align}
	\label{hamiltonian}
	\hat{h}_q = -\bm{\nabla}\cdot\frac{\hbar^2}{2m^*}\bm{\nabla}+U_q&
                    -i\bm{B}_q\cdot(\bm{\nabla}\times\bm{\sigma}) 
		    +\bm{\sigma}\cdot\bm{\Sigma}_q  \nonumber \\
		    &+\frac{1}{2i}(\bm{\nabla}\cdot\bm{I}_q+\bm{I}_q\cdot\bm{\nabla}).
\end{align}
For protons, one needs to add Coulomb potentials [Eqs.~(20) and (24) in Ref.~\cite{shi18}]. 
The detailed expression of $U_q$ can be found in Eq. (18) of Ref.~\cite{shi18}.  
The time-odd potentials included in
Eq.~(\ref{hamiltonian}) read
\begin{align}
	\label{time_odd_pot}
	\bm{\Sigma}_q&=\frac{1}{3}(-b_0+2b_0')\bm{s}-\frac{1}{3}(2b_0-b_0')\bm{s}_q \nonumber \\
		     &~~~~ ~~~~~~~~~~ ~~~~~~~~      -b_4\bm{\nabla}\times\bm{j}-b_4'\bm{\nabla}\times\bm{j}_q,\\
	\bm{I}_q     &=-2b_1\bm{j}+2b_1'\bm{j}_q-b_4\bm{\nabla}\times\bm{s}-b_4'\bm{\nabla}\times\bm{s}_q.
\end{align}

In the current work, those terms containing $\Delta\bm{s}$ are ignored. This is
because, frequently, the inclusion of the $\Delta\bm{s}$ and
$\bm{\nabla}\cdot\bm{s}$ terms is known to induce the finite-size
instabilities~\cite{hell12}. Hence, it has been the common practice for the
time-dependent applications of the DFT to ignore these terms, see
Ref.~\cite{stev16} for instance.

\subsection{Time propagation}

The nuclear non-relativistic time-dependent Schr\"odinger equation reads
\begin{equation}
	\label{tdhf}
	i\hbar\pdv{\psi_{i,q}(t)}{t}=\hat{h}_q(t)\psi_{i,q}(t),
\end{equation}
where $\hat{h}_q$ can be found in Eq.~(\ref{hamiltonian}). In this section, the
subscript $q$ is ignored for simplicity. The equation (\ref{tdhf}) has the
formal solution
\begin{equation}
	\psi_i(t)=\hat{\mathscr{U}}(t)\psi_i(0)=\hat{T}\exp(-\frac{i}{\hbar}\int_0^t\hat{h}(t')\,dt')\psi_i(0),
\end{equation}
where $\hat{\mathscr{U}}$ is the time-evolution operator, and $\hat{T}$ is the
time-ordering operator. To solve the time-dependent problem, one breaks up the
total time evolution into $N$ small increments of time $\Delta t$
\begin{equation}
	\hat{U}(t,t+\Delta t)=\exp(-\frac{i}{\hbar}\int_t^{t+\Delta t}\hat{h}(t')\,dt').
\end{equation}

The time-evolution operator $\hat{\mathscr{U}}(t)$ can be obtained by
consecutive actions of $\hat{U}(t,t+\Delta t)$
\begin{equation}
	\hat{\mathscr{U}}(t)=\prod_{n=0}^{N-1}\hat{U}(n\Delta t,(n+1)\Delta t).
\end{equation}

For small $\Delta t$ one could approximate $\hat{U}(t,t+\Delta t)$ by Taylor
expansion up to order $m$:
\begin{equation}
\label{expansion}
	\exp(-\frac{i}{\hbar}\hat{h}\Delta t)\approx\sum_{n=0}^m\frac{1}{n!}\left(\frac{-i\Delta t}{\hbar}\right)^n \hat{h}^n,
\end{equation}
where $\hat{h}$ has been assumed to be time independent in the time interval of
$\Delta t$. In the current work, $\Delta t$ is taken to be 0.2\,fm/$c$, and
$m=4$. These choices are motivated by previous TDHF calculations.

In the realistic calculations, each time advance of single-particle wave
functions $\psi_i$, from time $t$ to $t+\Delta t$, has been achieved by using
the Crank-Nicolson method~\cite{Bon76a}. Specifically, from a series of
single-particle wave functions at $t$, $\psi_i(t)$, one first performs
\begin{equation}
\label{time_progress}
\psi_i^{\rm temp}(t+\Delta t)=\hat{U}^{\rm t}(t,t+\Delta t) \psi_i(t). 
\end{equation}
Having $\psi_i^{\rm temp}(t+\Delta t)$, and $\psi_i(t)$, one assembles various
densities using respective single-particle wave functions, obtaining the
$\rho^{\rm temp}(t+\Delta t)$ and $\rho(t)$.

Using these densities, one obtains the densities at a ``middle time'',
$\rho^{\rm mid}(t+\frac{\Delta t}{2})=0.5[\rho^{\rm temp}(t+\Delta
t)+\rho(t)]$. Now, one constructs the Hamiltonian $\hat{h}^{\rm mid}$, using
$\rho^{\rm mid}(t+\frac{\Delta t}{2})$ [see Eq.~(15) of Ref.~\cite{shi18}, and
Eq.~(\ref{hamiltonian}) for the form of the Hamiltonian]. A second time
propagation operation $\hat{U}^{\rm mid}(t,t+\Delta t)$ with $\hat{h}^{\rm
mid}$ [Eq.~(\ref{expansion})] is performed on the single-particle levels,
finally obtaining the wave functions at $t+\Delta t$ 
\begin{equation}
\psi_i(t+\Delta t)=\hat{U}^{\rm mid}(t,t+\Delta t)\psi_i(t).  
\end{equation}
Here, $\hat{U}^{\rm mid}$ differs from $\hat{U}^{\rm t}$
[Eq.~(\ref{time_progress})] in that the former uses the single-particle
Hamiltonian in its exponent [Eq.~(\ref{expansion})] at the time $t+\frac{\Delta
t}{2}$, whereas the latter refers to the operator $\hat{U}$, where the
Hamiltonian is constructed using the quantities at the time $t$.

Note that in the above procedure, one has to perform the time propagation
twice. The single-particle Hamiltonian does not contain time specifically. In
realistic calculations, the unitarity of the operator
$\exp(-\frac{i}{\hbar}\hat{h}\Delta t)$ needs to be checked as it is
approximated using a Taylor expansion~[Eq.~(\ref{expansion})]. For the chosen
parameter, $\Delta t=0.2$\,fm/$c$ and $m=4$, we evaluate the matrix elements
\begin{equation}
\mathcal{I}_{ij} \equiv \langle \psi_i(t)|\hat{U}|\psi_j(t)\rangle 
                 \approx \langle \psi_i(t)|\psi_j(t+\Delta t)\rangle.
\end{equation}
Both the diagonal and off-diagonal matrix elements start to deviate from 1 and
0, respectively, at or after the 6th place after the decimal point. For a
better approximation of the $\hat{U}$ operator, one could decrease $\Delta t$
and increase $m$.

When the BCS pairing is included, the occupation amplitudes, $v_{i,q}$'s in
Eq.~(\ref{occupation}), are kept unchanged when calculating the densities
during the time development~\cite{maru14,Schuetrumpf2018}. When evaluating the
densities, the single-particle wave functions vary according to
Eq.~(\ref{tdhf}). This is a coarse approximation of dynamical pairing, as the
occupation probabilities should vary with time. Indeed, some of the problems
associated with the TDHF + BCS method in describing particle transport has been
discussed in Ref.~\cite{scamps12}. This approximation of the pairing will be
improved in our future publications. A natural extension would be to solve the
full time-dependent HFB problem~\cite{stetcu11,bulgac16,magi17}. Since the HFB
theory treats nuclear interactions in the particle-hole and pairing channels in
one single variational process~\cite{ring80}, a time-dependent HFB treatment
allows for the occupation amplitudes being determined dynamically by the upper
and lower components at a given time.

\subsection{Absorbing boundary conditions (ABC)}
\label{ABC}

With Dirichlet boundary conditions, it has been known that the TDDFT
calculations show the occurrence of non-physical particle densities at the
boundary region. To cure this problem, it has been proposed~\cite{naka05} to
use the so-called absorbing boundary conditions. This is achieved by
introducing an imaginary potential
\begin{subequations}
\label{eq:ABC}
\begin{equation}
\hat{h}(\vec{r})\rightarrow \hat{h}(\vec{r})+i\tilde{\eta}(\vec{r})
\end{equation}
at the boundary region of the form
\begin{equation}
\tilde{\eta}(\vec{r})=\left\{\begin{array}{ll}
0 & \mbox{for }0<|\vec{r}| \le R\\
\eta_0\frac{|\vec{r}|-R}{\Delta r} & \mbox{for } R<|\vec{r}|<R+\Delta r
\end{array}\right.\, .
\end{equation}
\end{subequations}
Recently, there have been efforts using more involved boundary
conditions~\cite{schu15,he19}. Based on these studies, we decide to use the ABC
due to its simplicity and effectiveness.

\subsection{IVD resonance calculations}
\label{gdr_cal}

The IVD resonance is the most common vibrational mode in nuclear physics, where
neutrons and protons vibrate against each other. This mode is responsible for
the $E1$ resonant strengths in the energy range of $\sim$10$-$20 MeV. This
broad peak is called giant dipole resonance (GDR)~\cite{hara01}. The current
work aims at a description of the IVD resonance in terms of the TDDFT in its
linearized limit, which is equivalent to the random-phase approximation
(RPA)~\cite{ring80}.

In the TDDFT description, the strength of this IVD vibrational mode can be
obtained by applying the following small boost on the obtained single-particle
wave functions, 
\begin{equation}
\label{boost}
\psi_{i,q}(\bm{r},\sigma;t=0+)\equiv\exp\left[-i\epsilon \sum_{\mu=-1}^{+1}\mathcal{M}(E1,\mu)\right]\psi_{i,q}(\bm{r},\sigma),
\end{equation}
with the IV operator $\mathcal{M}(E1,\mu)$ defined as 
\begin{equation}
	\label{ISV_operator}
	\mathcal{M}(E1,\mu)=e^{(E1)}_q r Y_{1\mu}(\vu{r})~~~~~~\mu=0,\pm 1,
\end{equation}
where $e^{(E1)}_p=Ne/A$, and $e^{(E1)}_n=-Ze/A$. When $\mathcal{M}(E1,\mu)$ acts
on neutron/proton single-particle wave functions, its coefficient takes value of 
$e^{(E1)}_p$/$e^{(E1)}_n$. The real spherical harmonics are defined as
\begin{equation}
\{Y_{1\mu}\}_{\mu=-1,0,1}=\{\sqrt{\frac{3}{4\pi}}\frac{\lambda}{r}\}_{\lambda=y,z,x}.
\end{equation}
In Eq.~(\ref{boost}), the boosted single-particle wave functions differ from
the static ones by including ``$t=0+$'', indicating their time-dependency. This
IV boost has to be small enough to ensure that the vibration is still within
the linearized regime. The typical magnitude of $|\epsilon|$ is
10$^{-3}$\,($e$\,fm)$^{-1}$. In this work, we apply 3D boost which has been
indicated by the summation over $\mu$ in the exponent in Eq.~(\ref{boost}).
For non-spherical nuclei, the 3D boost allows for obtaining the moments
along the three axes in a single run.
The boost is applied over the whole box, although a masking procedure works
better confining its effect in the range of the nucleus~\cite{paul20}.

The time evolution of the dipole moment 
\begin{equation}
\expval{\mathcal{M}(E1,\mu)} \equiv \int e^{(E1)}_n \rho_n r Y_{1\mu} d \bm{r}
                                  + \int e^{(E1)}_p \rho_p r Y_{1\mu} d \bm{r}
\end{equation}
is then recorded to certain length of time. Note, that although the boost is 3D,
the resulted $\expval{\mathcal{M}(E1,\mu)}$ are extracted for each $K$ component without summing them. 
The strengths are the Fourier transform of
$\langle\mathcal{M}(E1,\mu)\rangle(t)$
\begin{equation}
	\label{strengths}
	S(E;E1)=
	-\frac{1}{\pi \hbar \epsilon}{\rm Im}\sum_{\mu=-1}^{+1}\int \expval{\mathcal{M}(E1,\mu)}(t)\,dt\,e^{(iE-\Gamma/2) t/\hbar},
\end{equation}
where $\Gamma$ is a smoothing parameter. The photoabsorption cross section
associated with the IVD resonance is obtained as
follows~\cite{ring80}
\begin{equation}
        \label{cross_section}
	\sigma_\mathrm{abs.}=\frac{16\pi^3}{9\hbar c} E \times S(E;E1).
\end{equation}

For a nucleus localized in space, the translational symmetry is spontaneously
broken. This results in the existence of the spurious excitation of
center-of-mass modes in the self-consistent calculations. For a perfect IVD
vibrational calculation, however, these spurious modes cannot be excited as
they are completely decoupled with the IVD mode. In realistic calculations, due
to the finite size of the basis one is working, the spurious mode may appear at
finite excitation energy and may be mixed among the physical IVD excitations.
The current work based on a finite-difference representation for the TDDFT +
BCS code and the HO basis for the FAM QRPA code uses the IVD modes and we do
not see a prominent spurious peak with this operator. Thus in this work, the
mixture of the spurious modes in the IVD modes is expected to be small.

\subsection{Calculation of energy-weighted sum rule (EWSR) for the IVD vibration}

Another important aspect of the vibration calculations is the evaluation of
EWSR~\cite{ring80}, which is a useful check of the implementation of the TDDFT
code. In the TDDFT code, the sum rule is calculated using 
\begin{equation}
	\label{sum_rule_tdhf}
	m_1 = \int E \times S(E;E1) d E.
\end{equation}
Recently, the EWSR for the density functional theory has been systematically
derived in Refs.~\cite{hino15,hino19}. For the current IVD operator, the sum
rule using Eq.~(98) of Ref.~\cite{hino19} can be adapted as follows
\begin{align}
	\label{sum_rule}
	&m_1 = \sum_{\mu=-1}^{+1} \int d\bm{r}\big[\grad{(rY_{1\mu})}\big]^2
	\Big\{\frac{\hbar^2}{2m} \Big[{e^{(E1)}_n}^2 \rho_n \nonumber + {e^{(E1)}_p}^2\rho_p\Big]\\ 
        &+ (C_0^{\tau}-C_1^{\tau})\Big(e^{(E1)}_n+e^{(E1)}_p\Big)^2\rho_n\rho_p \nonumber \\
	&+ \sum_{k=0,1}(C_k^{\tau}+C_k^j) \Big[e^{(E1)}_n\rho_n + (-1)^{k+1}e^{(E1)}_p\rho_p\Big]^2\Big\}.
\end{align}
The definition of the spherical harmonics
can be found in Eq.~(\ref{ISV_operator}). The coupling constants in terms of
$C_{0,1}^{\tau, j}$ are related to $b_1,b_1'$ through
\begin{align}
C_0^{\tau}&=-C_0^j=b_1-0.5b_1', \\
C_1^{\tau}&=-C_1^j=-0.5b_1'.
\end{align}
If we define the kinetic-energy contribution
\begin{align}
\label{TRK}
m_1^{\rm kin} &= \sum_{\mu=-1}^{+1} \int d\bm{r}[\grad{(rY_{1\mu})}]^2 \frac{\hbar^2}{2m} \Big[{e^{(E1)}_n}^2 \rho_n + {e^{(E1)}_p}^2\rho_p\Big] \nonumber \\
              &=\frac{9}{4\pi}\frac{\hbar^2}{2m}\frac{NZ}{A}e^2,
\end{align}
then the enhancement factor, $\kappa$, due to the contribution of
interaction-energy term with respect to the kinetic part, can be calculated
through
\begin{equation}
\label{kappa}
m_1=m_1^{\rm kin}(1+\kappa).
\end{equation}
The classical sum rule of the IVD operator, which is the Thomas-Reiche-Kuhn
(TRK) sum rule~\cite{bohr75} can be analytically expressed as shown in
Eq.~(\ref{TRK}).

The EWSR value obtained from Eq.~(\ref{sum_rule}) are related to various
densities of the ground state. Thus, they can be determined rather precisely.
To what extent the $m_1$ values obtained from TDDFT [Eq.~(\ref{sum_rule_tdhf})]
and Eq.~(\ref{sum_rule}) agree, forms a stringent testing ground for the TDDFT
code.

\subsection{Sky3D calculations}
\label{sky3d}

To demonstrate the precision of the current code, it is necessary to benchmark
it against an existing code with an identical calculation. In this work, this
benchmark is done with a well established code Sky3D.

We use the Sky3D code as described in Refs.~\cite{maru14,Schuetrumpf2018}. An
important difference to the implementation presented in this code is that
derivatives are performed utilizing the fast Fourier transform and thus the
natural boundary conditions are periodic boundary conditions. The difference is
of special importance for time-dependent calculations, as it affects the
quantization of unbound energy states. Furthermore, when evaporated material is
leaving the box it is again introduced from the other side of the box and not
reflected as with Dirichlet boundary conditions. The codes differ slightly in
the way the density at middle time is approximated. In Sky3D the wave functions
are propagated until middle time $t+\Delta t/2$. These densities are then
directly taken to calculate the Hamiltonian at middle time
$\hat{h}^\mathrm{mid}$.

For the benchmarks we implemented the same boost as described in
Sec.~\ref{gdr_cal} and also the imaginary potential for the ABC from
Sec.~\ref{ABC}.

\section{results and discussions}
\label{results}

To complete the benchmark of the implemented TDDFT code, one has to include
careful calculations and compare the calculated results with those of existing
codes. Particularly useful testing cases for the TDDFT code are the
calculations of IVD resonance for light spherical and deformed nuclei.

In Ref.~\cite{naka05} careful comparative study has been done between the TDDFT
code and the RPA calculations. Detailed dipole-moment response as a function of
time, as well as the corresponding strengths results for $^{16}$O nucleus has
been presented with the specific force being provided. In this section, we
first present results of the current code, comparing them with those of Sky3D
code and Ref.~\cite{naka05}. The calculation is then extended to a spherical
nucleus $^{40}$Ca, as well as deformed magnesium isotopes $^{24,34}$Mg with
conventional Skyrme EDF SkM*~\cite{bart82}, and a more recent EDF
\UNEDFone~\cite{kort12}.

The \UNEDFone~EDF contains Lipkin-Nogami (LN) pairing~\cite{kort12} in the
parameter adjustment process. In principle, one has to include this part
specifically. However, we decide to be more flexible in the pairing treatment
for the current TDDFT calculations based on the two following considerations.
First, the original \UNEDFone~parameter is determined in the HO basis and with
specific cut-off on the HFB problem. Whereas the current code is working in the
3D Cartesian coordinate space. Hence, the continuum is discretized differently
from that of a HO code. Consequently, there is no way to make the pairing
treatment identical in the two codes~\cite{shi18}. Second, the observables we
are interested in, namely, the strengths for the IVD resonances are well known
to be insensitive to the pairing interactions~\cite{piek06}. The strengths
corresponding to pygmy dipole resonance (PDR) are only enhanced very marginally
by including the pairing interaction, as will be shown in Sec.~\ref{magnesium}.

\subsection{Results for light nuclei}
\label{light}

\subsubsection{Benchmark calculations for $^{16}$O with Skyrme SIII EDF}
\label{siii}

\begin{figure}
\centering
\includegraphics[scale=0.4]{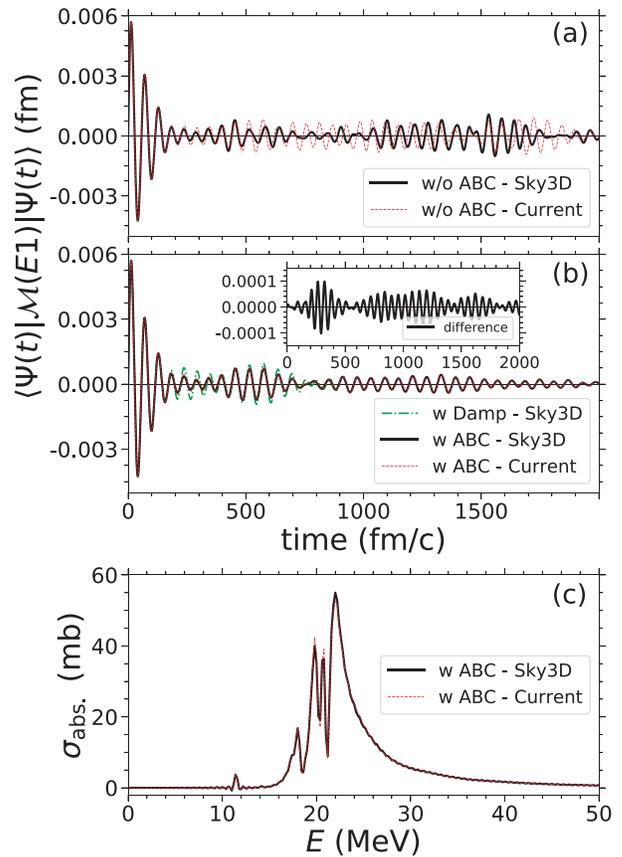}
\caption{The response functions, $\expval{\mathcal{M}(E1,\mu=0)}(t)$, of $^{16}$O
calculated with the current TDDFT code and the Sky3D code~\cite{maru14}. The
inset of panel (b) shows the difference between the $\expval{\mathcal{M}(E1,\mu=0)}$ 
values calcualted with the TDDFT and the Sky3D codes, both including the ABC.
The panel (c) shows the cross sections resulted from the respective response functions
in the panel (b). The smoothing parameter is $\Gamma=0$.}
\label{figure1}
\end{figure}

The nucleus $^{16}$O is of particular interest in theoretical benchmarking
calculations, as the structure of the strength is sensitive to the included
terms in the EDF~\cite{naka05}. Hence, many theoretical
methods~\cite{inak09,frac12,wu18} took $^{16}$O as a testing case for the
proposed method. In this section, we perform TDDFT calculations with Skyrme
force parameter SIII~\cite{liu76} with time-odd potentials in the form of
Eq.~(\ref{time_odd_pot}) (SIII-full), as well as SIII without any time-odd
contributions (SIII-even). 
For the calculations presented in Fig~\ref{figure1}, the time-odd 
potentials are identical with that of the Sky3D code
[Eqs. (8e) and (8f) of Ref.~\cite{maru14}]. Specifically, the time-odd potentials are
the same as SIII-full, except that the terms including {\it only} $\vec{s}$ are left out.

\begin{figure}
\centering
\includegraphics[scale=0.4]{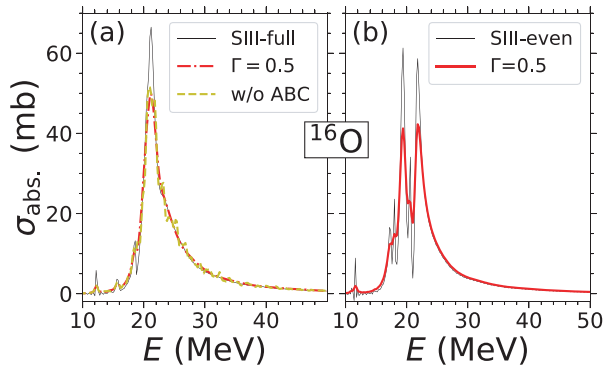}
	\caption{The calculated photoabsorption cross sections of $^{16}$O
using Eq.~(\ref{cross_section}). The left panel shows the results with
SIII-full EDF, whereas the right panel shows those with SIII-even. The thinner
black lines indicate the results without smoothing procedure.}
\label{figure2}
\end{figure}

Figure~\ref{figure1} displays a set of comparisons of responses of the dipole
moments between the currently implemented code and the Sky3D code.
Figure~\ref{figure1}(a) compares the response functions without any absorbing
mechanism. We see that the magnitude agrees well for $t\le400$\,fm/$c$.
However, the good agreement starts to deteriorate after $t\approx500$\,fm/$c$.
This is due to the different boundary conditions used in the two codes, which
results in the different treatment of the particle densities bounced back from
the border of the box. Indeed, even within the same code, using a finer grid
results in rather different response functions after certain time.

Figure~\ref{figure1}(b) compares the response functions with the ABC
[Eq.~(\ref{eq:ABC})] calculated with both codes. For both codes we use
$\eta_0=10\,\mathrm{MeV}$, $R=10\,\mathrm{fm}$, and $\Delta r=12\,\mathrm{fm}$.
It can be seen that with the same ABC, both codes give almost identical
response functions. The difference of the dipole moment, shown in the inset, is
at least an order of magnitude smaller than the original moment value.
Fig.~\ref{figure1}(c) shows the photoabsorption cross sections calculated from
the response functions shown in Fig.~\ref{figure1}(b). Again, the agreement is
remarkable.

Figure~\ref{figure2} shows the calculated photoabsorption cross sections for
SIII-full and SIII-even, with $\Gamma$=0.5\,MeV and without the smoothing
procedure ($\Gamma$=0). Again, a good correspondence can be seen between
Fig.~6(b) of Ref.~\cite{naka05} and Fig.~\ref{figure2} of the current work.
Specifically, for SIII-full we see, for both results, that the single largest
peak occurs at $E\approx21.2$\,MeV. For SIII-even, the two peaks occur at
$E\approx19.4$ and 21.8\,MeV for both the current result and those shown in
Fig.~6(b) of Ref.~\cite{naka05}. Using a smoothing parameter of
$\Gamma=0.5$\,MeV brings the general energy dependence of the photoabsorption
cross sections rather close to those given in Fig.~6(b) of Ref.~\cite{naka05}.

In these calculations, we use the ABC as described in Sec.~\ref{ABC}. In
Fig.~\ref{figure2}(a), we also include the results without the ABC. It can be
seen that the strength without the ABC differs from that with the ABC in that
the former gives small peaks for excitation energies larger than that
corresponds to the main peak. These small peaks are spurious which are removed
by absorbing potential in the outer layer region.

\subsubsection{Comparing TDDFT with FAM-RPA: $^{16}$O and $^{40}$Ca}
\label{spherical}

In this section, we compare our TDDFT approach to the RPA calculation based on
the linear-response formalism, the finite-amplitude method
(FAM)~\cite{naka07,avog11}. The FAM allows us to calculate the response
function without constructing the QRPA matrices in the case of the nuclear DFT.
The present implementation of the FAM-QRPA~\cite{kort15} is based on the
nuclear DFT solver \HFBTHO~\cite{stoi05,stoi13,perez17}, which allows to
describe the superconducting axially deformed nuclei in the HO basis.

Before showing the cross-section results, we first present the calculated
static properties using both codes. Table~\ref{table1} lists the calculated
ground-state energy decomposition into various terms, as well as the
root-mean-square radii. For a fixed box size, three different grid spacings
have been used. It can be seen that the ground-state energy is overbound by
$<$200\,keV using the coarsest grid with $dx=1.0$\,fm. Using finer grid
spacings reduces the total energy differences to $\le$50\,keV. It should be
noted that, the seemingly poor accuracy of a spacing of 1.0 fm does not
drastically affect the dynamic calculation [see Fig.~\ref{figure1}(b)].

\begin{table*}[htb]
	\caption{The calculated static properties for $^{16}$O 
and $^{40}$Ca with \UNEDFone~EDF, using the current TDDFT and the \HFBTHO~codes. 
For the TDDFT calculations, the simulation boxes have dimensions 
of [$-$14.5,+14.5]$^3$ fm$^3$. Three grid spacings have been
used to see the convergence of the TDDFT calculations. For the
\HFBTHO~calculations, 20 HO shells are used.}
\label{table1}
\begin{ruledtabular}
\begin{tabular}{lrrrrrrrr}
        & \multicolumn{4}{c}{$^{16}$O} & \multicolumn{4}{c}{$^{40}$Ca}  \\
        \cline{2-5} \cline{6-9}
	& \multicolumn{3}{c}{Current} & \multirow{2}{*}{\HFBTHO} & \multicolumn{3}{c}{Current}& \multirow{2}{*}{\HFBTHO} \\
        \cline{2-4} \cline{6-8}
        & $\Delta x$=1.0 fm &$\Delta x$=0.784 fm & $\Delta x$=0.707 fm & &  $\Delta x$=1.0 fm & $\Delta x$=0.784 fm & $\Delta x$=0.707 fm & \\
\hline
	$E_{\rm tot}$ (MeV)               & $-$121.139   & $-$120.997& $-$120.986 & $-$121.000  & $-$340.873  & $-$340.599  & $-$340.571 & $-$340.625 \\
	$E_{\rm Kin.}$ (MeV)              & 236.905      & 236.443   & 236.414    & 236.494     & 659.414     & 658.387     & 658.290    & 658.505 \\
	$E_{\rho}$ (MeV)                  & $-$406.666   & $-$405.978& $-$405.936 & $-$406.055  & $-$1137.525 & $-$1135.918 & $-$1135.749& $-$1136.071 \\
	$E_{\tau}$ (MeV)                  & $-$0.890     & $-$0.886  & $-$0.886   & $-$0.886    & $-$3.218    & $-$3.209    & $-$3.207   & $-$3.209 \\
	$E_{\Delta\rho}$ (MeV)            & 36.522       & 36.486    & 36.492     & 36.520      & 68.788      &68.601       & 68.575     & 68.640 \\
	$E_{\rm SO}$ (MeV)                & $-$0.636     & $-$0.671  & $-$0.677   & $-$0.681    & $-$0.979    & $-$1.046    & $-$1.059   & $-$1.077 \\
	$E_{\rm dir.}^{\rm Coul.}$ (MeV)  & 16.448       & 16.429    & 16.427     & 16.428      & 80.201      & 80.132      & 80.124     & 80.134 \\
	$E_{\rm exc.}^{\rm Coul.}$ (MeV)  & $-$2.823     & $-$2.820  & $-$2.820   & $-$2.820    & $-$7.554    & $-$7.547    & $-$7.546   & $-$7.548 \\
	$r_{\rm rms}^{\nu}$ (fm)          & 2.666        & 2.669     & 2.669      & 2.668       &  3.360      & 3.362       & 3.362      & 3.362 \\
	$r_{\rm rms}^{\pi}$ (fm)          & 2.684        & 2.687     & 2.687      & 2.686       &  3.395      & 3.398       & 3.398      & 3.398 \\
	$r_{\rm rms}^{\rm tot.}$ (fm)     & 2.675        & 2.678     & 2.678      & 2.677       &  3.377      & 3.380       & 3.380      & 3.380 \\
\end{tabular}
\end{ruledtabular}
\end{table*}

\begin{figure}
\centering
\includegraphics[scale=0.25]{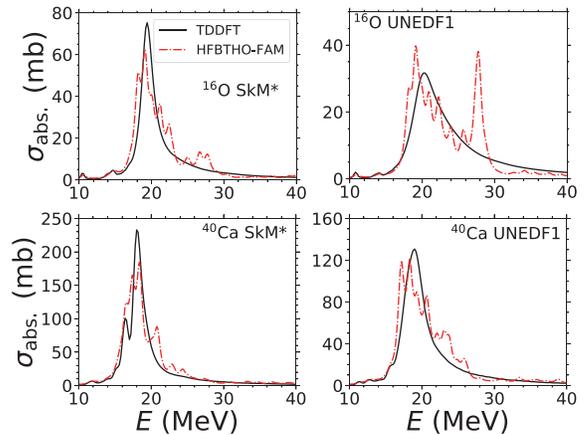} 
\caption{The calculated photoabsorption cross sections for $^{16}$O 
and $^{40}$Ca, using TDDFT of the present implementation and FAM-RPA 
based on the \HFBTHO~with the SkM* and \UNEDFone~EDFs.}
\label{figure3}
\end{figure}

Figure~\ref{figure3} shows the photoabsorption cross sections for $^{16}$O and
$^{40}$Ca calculated with the TDDFT and FAM-RPA. The energy of the main peak
and the low-energy side of the main peak agree well between the two approaches,
while the high-energy tail part is more fragmented in the FAM-RPA strength.
This behavior found in the calculation using the HO basis is also found in the
QRPA calculations in deformed nuclei using the HO basis~\cite{peru08,losa10}.

\begin{figure}
\centering
\includegraphics[scale=0.40]{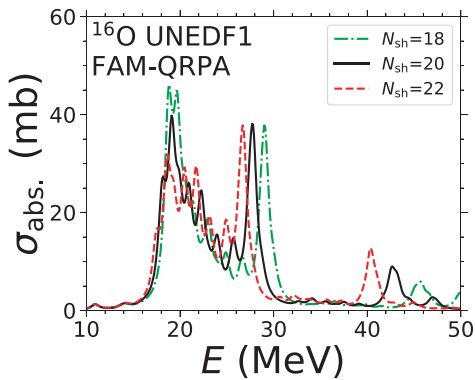} 
\caption{The calculated photoabsorption cross sections for $^{16}$O, using
FAM-RPA method based on the \HFBTHO~with the \UNEDFone~EDF. Three different HO basis numbers are used 
to see the convergence of the results.}
\label{figure4}
\end{figure}

Figure~\ref{figure4} displays the FAM-RPA results with increasing number of
HO basis ($N_{\rm sh}$). We see that the peak at $E\approx25-35$\,MeV moves toward
the main peak with increasing $N_{\rm sh}$. We note a slow convergence of the
strength function for this nucleus with this particular EDF. For medium-heavy
nuclei, the isoscalar and isovector multipole strength functions are found to
be converged already at $N_{\rm sh}=20$~\cite{stoit11,oishi16}.

\subsubsection{Results for deformed nuclei: $^{24,34}$Mg}
\label{magnesium}

\begin{table}[htb]
	\caption{Calculated static DFT results for $^{24}$Mg using SkM* and
\UNEDFone~EDFs. A comparison is made between the results using the current TDDFT and the \HFBTHO~codes~\cite{stoi05,stoi13}. There is no center-of-mass
correction for \UNEDFone~calculations. The quadrupole moments are defined as
$Q_{20}=2\expval{\hat{z}^2}-\expval{\hat{x}^2}-\expval{\hat{y}^2}$. The
single-particle levels are doubly degenerate and labeled by $\Omega^{\rm parity}$,
where $\Omega$ denotes the total angular momentum of the level projected onto
the $z$-axis. All quantities are in units of MeV, except for $Q_{20}$ values
which are in barn.}
\label{table2}
\begin{ruledtabular}
\begin{tabular}{lrrrr}
	& \multicolumn{2}{c}{SkM*} & \multicolumn{2}{c}{\UNEDFone} \\
         \cline{2-3} \cline{4-5}
        & Current   &  \HFBTHO  & Current  &  \HFBTHO   \\
       \hline
	$E_{\rm tot}$                   & $-$197.123  & $-$197.155 & $-$189.881  & $-$189.852        \\
	$E_{\rm Kin.+c.m.}$             & 384.483     & 384.091   & 401.148     &  400.387       \\
	$E_{\rm Coul}$                  & 28.681      & 28.650    & 28.713      &  28.671       \\
	$E_{\rm Skyrme}$                & $-$610.287  & $-$609.896 & $-$619.742  & $-$618.910        \\
	$Q_{20}$                        & 1.072       & 1.072    &  1.126        & 1.137        \\
	$\epsilon^{\pi}_{1/2^+}$        & $-$34.236   & $-$34.249    & $-$29.474   &  $-$29.480       \\
	$\epsilon^{\pi}_{1/2^-}$        & $-$23.510   & $-$23.528    &  $-$20.865  &  $-$20.898       \\
	$\epsilon^{\pi}_{3/2^-}$        & $-$19.429   & $-$19.396   & $-$17.348   &   $-$17.285      \\
	$\epsilon^{\pi}_{1/2^-}$        & $-$13.945   & $-$13.973    & $-$13.219   &  $-$13.196       \\
	$\epsilon^{\pi}_{1/2^+}$        & $-$12.066   & $-$12.075    & $-$11.036   &  $-$11.064       \\
	$\epsilon^{\pi}_{3/2^+}$        & $-$9.525    & $-$9.519   &  $-$8.596   &    $-$8.585     \\
	$\epsilon^{\nu}_{1/2^+}$        & $-$39.279   & $-$39.290    & $-$34.215   &  $-$34.218       \\
	$\epsilon^{\nu}_{1/2^-}$        & $-$28.361   & $-$28.377    &  $-$25.500  &  $-$25.529       \\
	$\epsilon^{\nu}_{3/2^-}$        & $-$24.274   & $-$24.235    & $-$22.034   &  $-$21.964       \\
	$\epsilon^{\nu}_{1/2^-}$        & $-$18.667   & $-$18.694    & $-$17.833   &  $-$17.806       \\
	$\epsilon^{\nu}_{1/2^+}$        & $-$16.725   & $-$16.729    & $-$15.604   &  $-$15.626       \\
	$\epsilon^{\nu}_{3/2^+}$        & $-$14.141   & $-$14.131    &  $-$13.148  &  $-$13.132       \\
\end{tabular}
\end{ruledtabular}
\end{table}

The nucleus $^{24}$Mg is one of the lightest nuclei with large prolate
deformation. Hence, the IVD vibration motion of this nucleus has been
frequently used as a testing case for TDDFT or RPA codes. Another interesting
system that has a prolately deformed ground state is $^{34}$Mg. The occurrence
of non-zero strength below $E=10$\,MeV in $^{34}$Mg is a signature of the pygmy
mode for neutron-rich Mg isotopes~\cite{ebat10}. For neutrons, there is a
pairing correlation which makes $^{34}$Mg particularly interesting. In this
section, we focus on the description of $^{24,34}$Mg with both TDDFT and the
FAM-QRPA methods.

\begin{figure}
\centering
\includegraphics[scale=0.55]{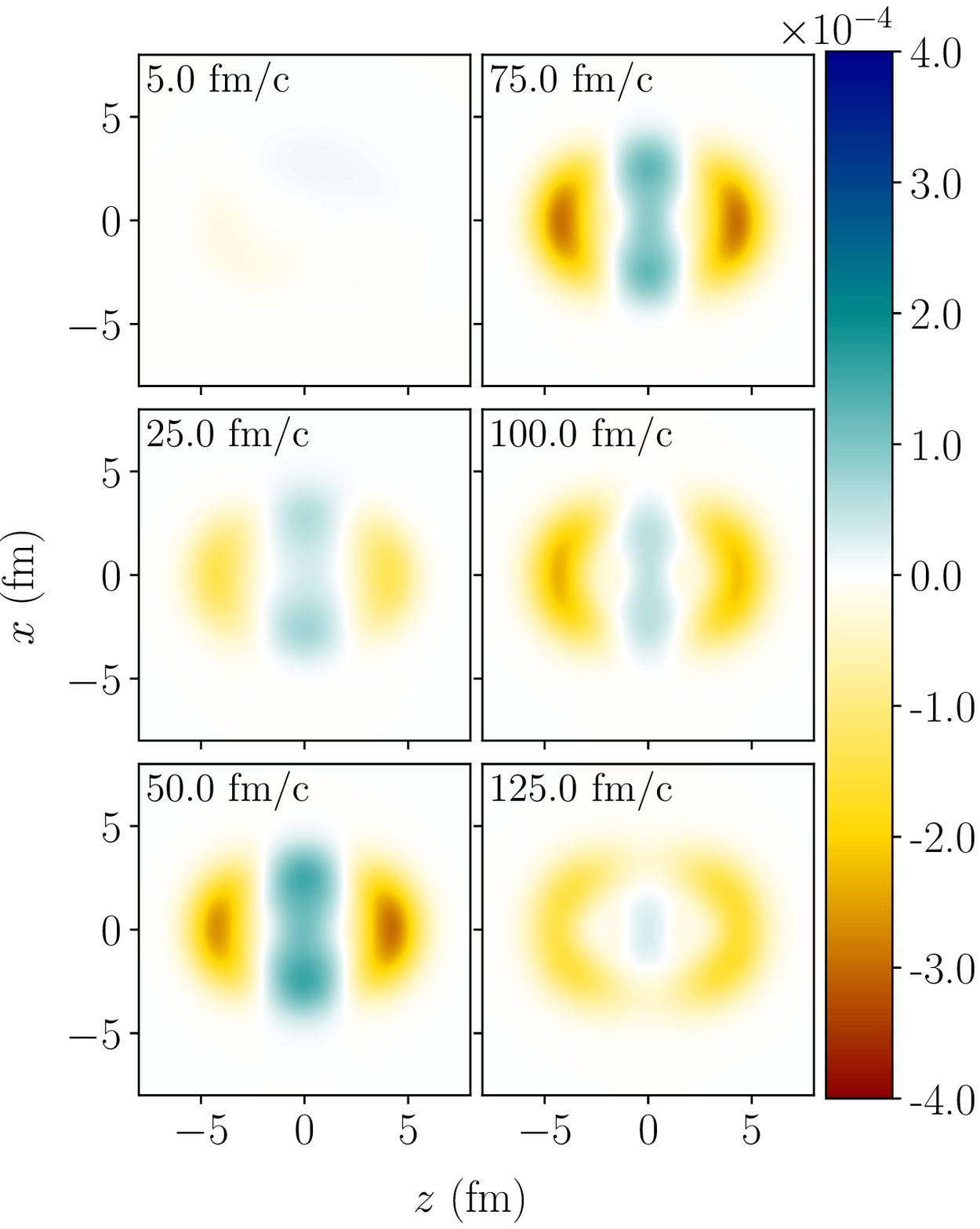}
	\caption{Time evolution of the IV density $\rho_p(\vb{r})- \rho_n(\vb{r})$
(in $\mathrm{fm}^{-3}$) in the $x-z$ plane ($y=0$) for the IVD mode in
$^{24}$Mg.}
\label{figure5}
\end{figure}

Table~\ref{table2} lists the calculated static information on $^{24}$Mg with
SkM* and \UNEDFone~EDFs. Figure~\ref{figure5} plots the IV densities,
$\rho_p(\vb{r})-\rho_n(\vb{r})$, on the $x-z$ plane with $y=0$, at a few
instances. As the neutrons and protons vibrate against each other, a fading and
strengthening pattern of the color can be seen. Careful examination reveals the
left-right and up-down asymmetry, which is due to the 3D boost that has been
initiated in the current calculations.

\begin{figure}
\centering
\includegraphics[scale=0.4]{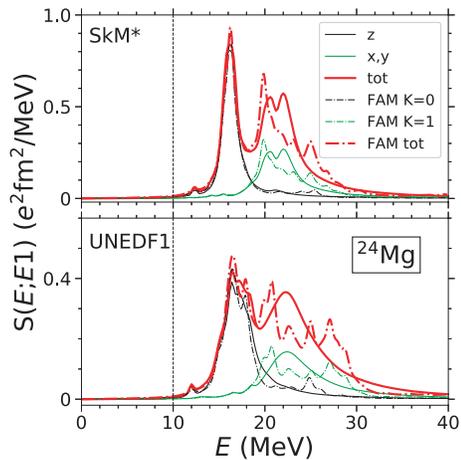}
	\caption{The calculated strength functions of $^{24}$Mg using SkM* and
\UNEDFone~EDFs, with the TDDFT and the FAM-RPA methods.}
\label{figure6}
\end{figure}

Figure~\ref{figure6} compares the strengths calculated with SkM* and
\UNEDFone~EDFs. It can be seen that the two peaks calculated with \UNEDFone~EDF
are considerably lower and broader compared to those calculated with SkM* EDF.
The positions of the two peaks are a few hundreds of keV higher for
\UNEDFone~EDF compared to those of SkM* EDF. In Fig.~\ref{figure6}, we plot our
FAM-RPA results too. The strength functions are almost identical up to the
first peak, after which the FAM-RPA calculations show more fragmented second
peak or sub-peaks compared to the TDDFT calculations. This spurious behavior of
the HO-basis calculation is similar to the TDDFT calculation without the ABC.
Both results indicate that the proper treatment of the boundary condition is
important to accurately describe the higher excitation energy region of the
strength distribution.

For the strength function of $^{24}$Mg calculated with SkM* EDF, there are a
few calculations using different models. For example, in Ref.~\cite{inak09},
the photoabsorption cross section for $^{24}$Mg has been calculated with the
FAM-RPA method. In Ref.~\cite{ebat10} a canonical-basis TDHFB calculation is
performed to calculate the $E1$ strength in $^{24}$Mg. In particular, the
result is consistent with their QRPA results~\cite{ebat10}. In
Ref.~\cite{losa10}, the QRPA calculations using (transformed) HO basis has been
performed for the $E1$ strengths in Mg isotopes.

Comparing these three existing results [Fig. 8(g) of Ref.~\cite{inak09}, Fig. 2
of Ref.~\cite{ebat10}, and Fig. 15 of Ref.~\cite{losa10}] with that in the
current work which is shown in Fig.~\ref{figure6}, it can be summarized that,
for all the calculated results, there are unambiguously two peaks at
$E\approx16$ and 22\,MeV. The structure or sub-peaks appearing between these
two are susceptible to, presumably, either the box size, or the truncation in
the single-particle levels, and HO shells used in the respective models. It is
rewarding to see such a consistency among independent methods and
implementations.

\begin{figure}
\centering
\includegraphics[scale=0.45]{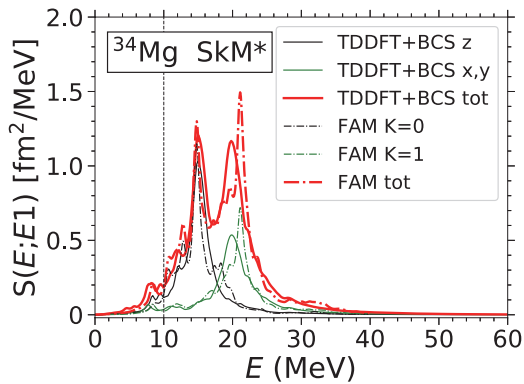}
	\caption{The calculated strength functions of $^{34}$Mg with SkM* EDF
using the TDDFT + BCS and the FAM-QRPA methods.}
\label{figure7}
\end{figure}

Figure~\ref{figure7} shows the calculated $E1$ strengths for $^{34}$Mg using
both the TDDFT + BCS and the FAM-QRPA calculations. For the TDDFT + BCS
calculations, pairing exists only for neutrons. The pairing strength for
neutrons is $V_n=-500$\,MeV\,fm$^3$. There are 44 single-neutron levels
included in the BCS problem. The highest-energy single-particle level has
$\epsilon=3.85$\,MeV.  To make the two methods comparable, we have fine tuned
the pairing strengths in the \HFBTHO~calculation in such a way that both codes
give similar pairing energies in the static calculations.

We see from Fig.~\ref{figure7} that both calculations yield two peaks at
$E\approx15$ and 20\,MeV.  Again, the second peak from the FAM-QRPA calculation
is slightly more fragmented compared to that from the TDDFT+BCS calculations.
These results are consistent with the canonical-basis TDHFB results of
Ref.~\cite{ebat10}.

For the neutron-rich oxygen, neon, and magnesium isotopes, the appearance of
the $E1$ strength below 10\,MeV are of particular
interest~\cite{cao05,ebat10,wang17}, as they correspond to the pygmy mode of
vibration. It has been shown \cite{ebat10} that the inclusion of the pairing
correlation would result in a small enhancement of the fraction of the
strengths below 10\,MeV, compared to a TDDFT result.

We compute the following PDR fraction~\cite{ebat10,ebata14}
\begin{equation}
\label{fraction}
	f_{\rm PDR}=\frac{m_1(E_c)}{m_1}\equiv\frac{\int^{E_c}E \times S(E)dE}{\int E \times S(E)dE},
\end{equation}
for the strength functions from TDDFT calculations with and without pairing.
The $f_{\rm PDR}$ value for $E_c=10$\,MeV is 2.3\% for the $E1$ strength
without pairing.  When the pairing is included, this quantity increases to
2.7\%, which is consistent with the results in Ref.~\cite{ebat10}.

\subsubsection{Calculated EWSR}

Table~\ref{table3} compares the $m_1$ values calculated with the ground-state
expectation value [Eq.~(\ref{sum_rule})], and those calculated with the
strength function obtained from the TDDFT method [Eq.~(\ref{sum_rule_tdhf})].
We see that the $m_1$ values from the TDDFT and those from Eq.~(\ref{sum_rule})
are rather close. The TDDFT values are systematically smaller than those of
Eq.~(\ref{sum_rule}) by less than 1\% of the $m_1$ values. This indicates the
correctness and good precision of the current implementation of the TDDFT code.

The classical TRK sum-rules [Eq.~(\ref{TRK})] are 59.2, 148.0, and 88.8
$e^2$\,fm$^2$\,MeV for $^{16}$O, $^{40}$Ca, and $^{24}$Mg, respectively. We
have computed the enhancement factor $\kappa$ using Eq.~(\ref{kappa}), which
are roughly 0.15 and 0.30 for each nucleus using \UNEDFone~and SkM* EDFs,
respectively.

\begin{table}[htb]
	\caption{The EWSR values of the IVD operator (in $e^2$\,fm$^2$\,MeV)
calculated using the current TDDFT code, compared to the ground-state values.
For the TDDFT results calculated with Eq.~(\ref{sum_rule_tdhf}), the integrations
are taken from 0 to 80\,MeV, with $\Gamma=0$ in
Eq.~(\ref{strengths}).}
\label{table3}
\begin{ruledtabular}
\begin{tabular}{lcc}
	& TDDFT & g.s. value\\
	\hline
	$^{16}$O (SIII-even)  &    67.1 &   67.3    \\
	$^{16}$O (SIII-full)  &    75.0 &   75.3    \\
	$^{16}$O (SkM*)       &    72.5 &   72.8    \\
	$^{16}$O (\UNEDFone)     &    67.0 &   67.6    \\
	$^{40}$Ca (SkM*)      &   194.0 &   194.9   \\
	$^{40}$Ca (\UNEDFone)    &   171.4 &   172.8   \\
	$^{24}$Mg (SkM*)      &   113.7 &   114.3   \\
	$^{24}$Mg (\UNEDFone)    &   101.8 &   102.9   \\
\end{tabular}
\end{ruledtabular}
\end{table}

\subsection{Results for Zr, Mo, and Ru nuclei}
\label{zr_mo_ru}

In the previous TDDFT + BCS calculations for light spherical and deformed
nuclei, we have seen the usefulness of the newly developed code. In this
section, we perform systematic calculations for the photoabsorption cross
sections of Zr, Mo, and Ru nuclei. For the TDDFT + BCS calculations, the box
size is $[-14.5,+14.5]^3$\,fm$^3$, with a uniform grid spacing of 1\,fm. In the
BCS pairing treatment, 100 and 70 neutron and proton single-particle wave
functions are included. The pairing strengths for neutrons and protons are
$V_n=-382$ and $V_p=-440$\,MeV\,fm$^3$, respectively. These are determined to
match the pairing energies of $^{106}$Mo using the above BCS setup, with those
given by the HFB results using the original \UNEDFone~EDF. 
The ABC has been always included with $\Delta r=16$\,fm and $\eta_0=10$\,MeV.
The above choice of the absorbing parameters seem to be effective for
excitation energies larger than 5.5\,MeV~\cite{naka05}, which is the energy
corresponding to the lower end of the GDR peaks.

In this section, we discuss the following related topics: the choice of
parameters used to describe the IV $E1$ cross sections of $^{92}$Mo
(Sec.~\ref{parameters}); the systematic TDDFT + BCS results for the spherical
Zr and Mo nuclei (Sec.~\ref{spherical}); the calculated potential-energy
surfaces for the ground states of neutron-rich Zr, Mo and Ru isotopes
(Sec.~\ref{static}); a case study of $^{100}$Mo in terms of the possible shape
coexistence (Sec.~\ref{mo100}); and the systematic predictions of the cross
sections for the heavier Zr (Sec.~\ref{deformed_curve}), Mo, and Ru
(Sec.~\ref{mo_ru}) isotopes, discussing the dynamical results in connection
with the shape coexistence and the evolution of triaxiality with neutron
number.

\subsubsection{The choice of parameters}
\label{parameters}

\begin{figure}
\centering
\includegraphics[scale=0.5]{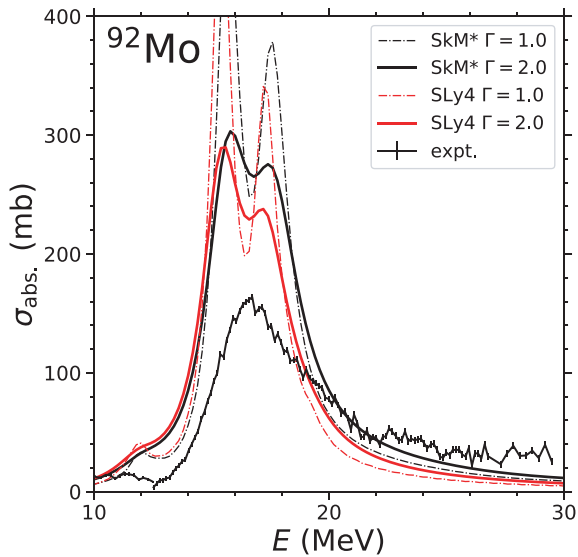}
	\caption{The IV $E1$ cross sections calculated using the SkM* and SLy4
EDFs with the smoothing parameters $\Gamma=1.0$ and  2.0\,MeV. The experimental
data are from Ref.~\cite{beil74}, which are extracted from
Refs.~\cite{exfor,nndc}.}
\label{figure89}
\end{figure}

Figure~\ref{figure89} compares the IV $E1$ cross sections of $^{92}$Mo
calculated using the SkM*~\cite{bart82} and SLy4~\cite{chab98} EDFs with the
experimental data~\cite{beil74}. The ground state of this semi-magic nucleus
($N=50$) is calculated to be spherical with various EDFs. For $E$ in the
interval of 14 and 20 MeV, we see pronounced strengths for both calculations,
as well as experimental data. The calculations with both SkM* and SLy4 show two
GDR peaks, which is at variance with the data which appears to have only one
peak. The RPA results of $^{92}$Mo calculated with SkM* in Ref.~\cite{kvasil09}
shows two peaks between 14 and 20 MeV, which is in agreement of the current
results. In Ref.~\cite{kvasil09} the second peak has a larger strength, whereas
in the current work the first one has a larger strength. Similar peak
structures are also seen in Ref.~\cite{inak09} for $^{90}$Zr, calculated with
SkM* EDF.

\begin{figure}
\centering
\includegraphics[scale=0.5]{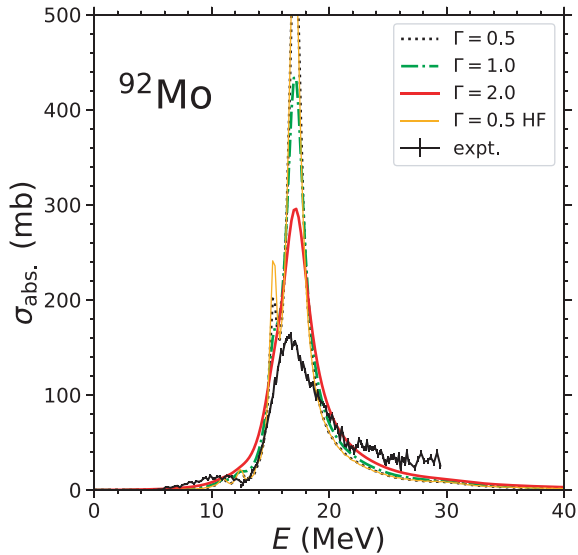}
	\caption{Similar to Fig.~\ref{figure89}, except that the calculations are performed with
\UNEDFone~EDF for $\Gamma=0.5$, 1.0, and 2.0\,MeV. A photoabsorption cross section calculated 
without pairing with $\Gamma=0.5$\,MeV MeV is also shown.}
\label{figure88}
\end{figure}

Figure~\ref{figure88} shows the calculated IV $E1$ cross sections of $^{92}$Mo
using \UNEDFone~EDF for $\Gamma=0.5$, 1.0, and 2.0\,MeV. We see that, as
expected, the centroids of the GDR peaks are the same. The photoabsorption
cross section calculated with smaller $\Gamma$ value is more concentrated
around the peak with a narrower energy width. Comparing the calculated results
with the cross-section data for the ($\gamma$,n)+($\gamma$,2n)+($\gamma$,3n)
reactions~\cite{beil74}, we see that the photoabsorption cross section
calculated with $\Gamma=2.0$\,MeV is stilll more concentrated around the peak
energy.

Although for $^{92}$Mo the calculated results with $\Gamma=2.0$\,MeV
overestimate the photoabsorption cross section, for the Zr isotopes, the
calculated heights of the GDR peaks are consistent with data using
\UNEDFone~EDF, as we will see later in Sec.~\ref{spherical}. Hence, we choose
to use $\Gamma=2.0$\,MeV for the remaining calculations in this section. This
choice of $\Gamma$ value is also consistent with the RPA
calculations~\cite{kvasil09}, where a 2.0\,MeV smoothing parameter was seen to
produce reasonable descriptions for these cross-section data.

As shown in Fig.~\ref{figure88}, with $\Gamma=2.0$\,MeV, the GDR curve
reproduces the rising part of the experimental data. It peaks at
$E\approx17.2$\,MeV and reproduces the experimental data of 17.13\,MeV. For the
lower part of the spectrum, the calculation with $\Gamma=2.0$\,MeV
underestimates the experimental cross section. For the result of
$\Gamma=0.5$\,MeV, the photoaborption cross section calculated without the
proton pairing is also plotted (the neutron pairing vanishes for this nucleus).
We see the photoabsorption cross sections calculated with and without the
proton pairing are almost identical except for the peaks where those of the
unpaired result are slightly higher. When a larger $\Gamma$ is used, the
results with and without the pairing interaction are even closer. Although the
inclusion of the pairing seems to have little influence on the IVD vibration
calculations, for deformed nuclei, it impacts the shapes of GDR peaks through
changing the deformations of the ground states.

\subsubsection{Results for spherical Zr and Mo isotopes}
\label{spherical}

\begin{figure*}
\centering
\includegraphics[scale=0.8]{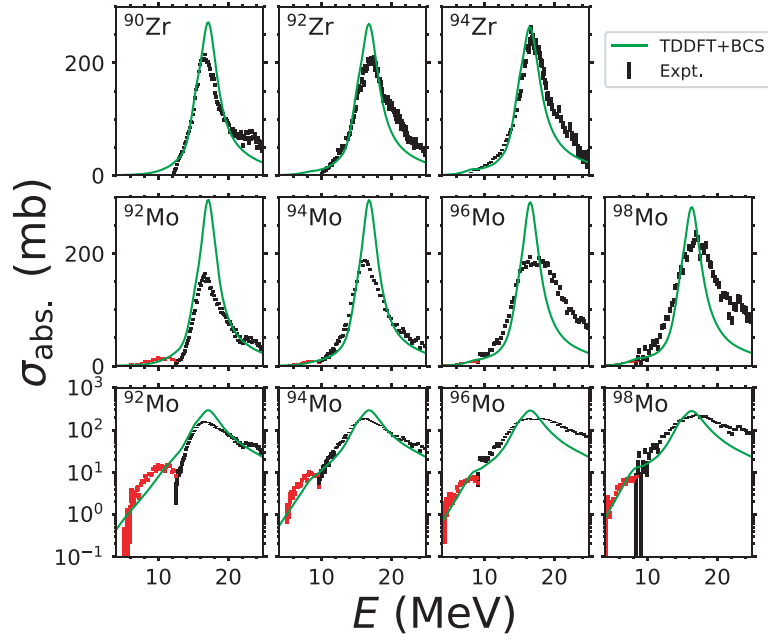}
	\caption{Calculated photoabsorption cross sections using \UNEDFone~EDF,
with smoothing parameter $\Gamma=2.0$\,MeV. The experimental data for GDR (red
and black crosses) are from Refs.~\cite{berman67,beil74}.  In
the third row, the same cross sections for Mo isotopes
are plotted in the logarithmic scale. The data for the
lower-energy parts (red crosses) are from
Refs.~\cite{rusev06,rusev08,rusev09,erhard10,utsu13}. The numbers are extracted from
Refs.~\cite{exfor,nndc}.}
\label{figure8}
\end{figure*}

In this section, we extend the calculation setup described in
Sec.~\ref{parameters} to calculate the remaining spherical Zr and Mo nuclei
where experimental data exist: $^{90,94}$Zr and $^{92,94,96,98}$Mo. The results
are shown in Fig.~\ref{figure8}. We see that the widths and the centroids of
the GDR peaks for $^{90,92,94}$Zr are well reproduced by the current
calculations. For $^{90,92}$Zr, the heights of the GDR peaks are overestimated.
For $^{92,94}$Mo, the centroids are slightly overestimated. The heights of the
GDR peaks of $^{92,94,96}$Mo are again overestimated. For $^{96,98}$Mo, we see
a flattening of the peaks in the experimental data, this might indicate the
triaxial deformations of the ground states, as will be discussed in the case of
$^{100}$Mo in Sec.~\ref{mo100}.

For the Mo isotopes, the low-energy part ($E\le10$\,MeV) of the cross sections
were observed using bremsstrahlung method~\cite{rusev06,rusev08,erhard10}. Our
calculated results reproduce the existence of the shoulders near $E=10$\,MeV.
For the low-energy part ($E<10$\,MeV), the calculated results underestimate the
cross-section data in $^{92,94}$Mo, and show reasonable agreement with
experimental data for $^{96,98}$Mo, as well as $^{100}$Mo, the latter of which
will be discussed in detail in Sec.~\ref{mo100}.

\subsubsection{Static potential energy surfaces for neutron-rich Zr, Mo, and Ru
\label{static}
isotopes}

\begin{figure*}
\centering
\includegraphics[scale=0.340]{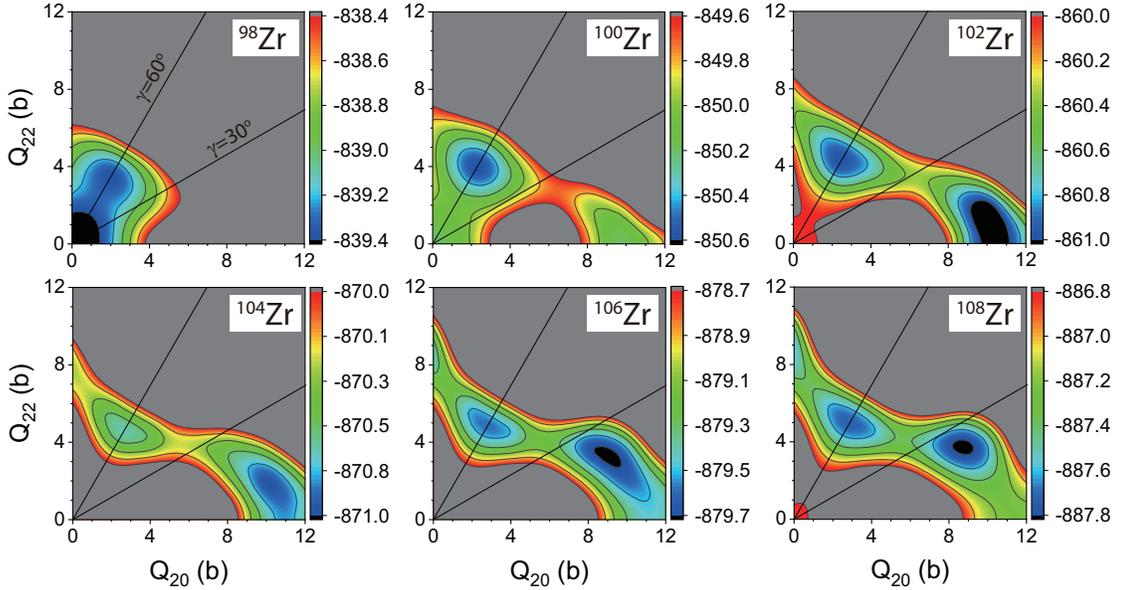}
	\caption{Calculated potential-energy surfaces for $^{98-108}$Zr, using
the \UNEDFone~EDF. The energies are in MeV.}
\label{figure9}
\end{figure*}

\begin{figure*}
\centering
\includegraphics[scale=0.34]{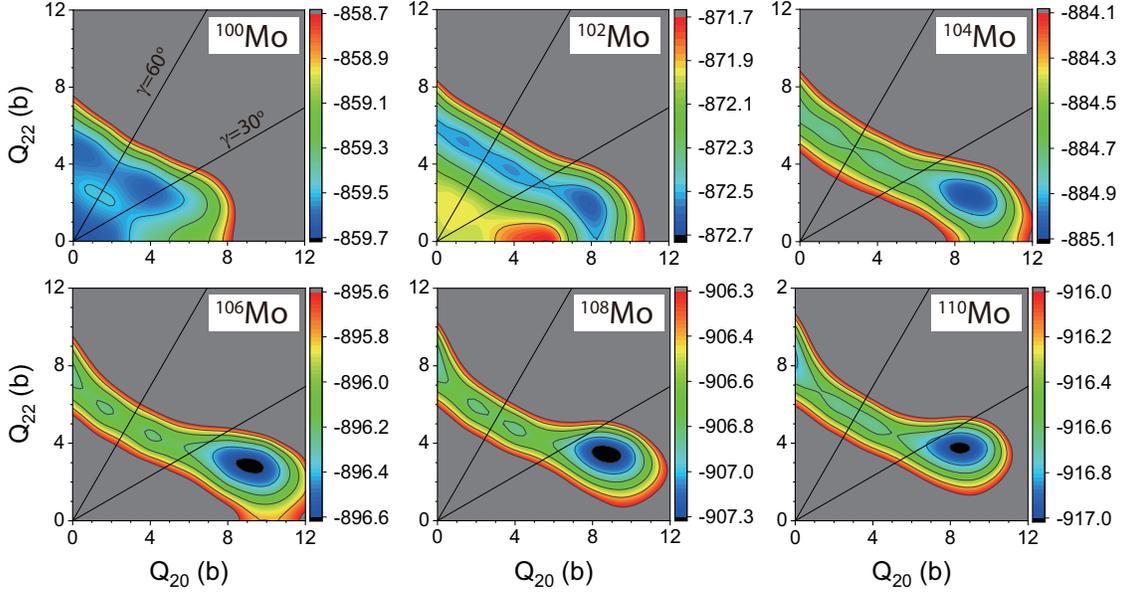}
	\caption{The same as Fig.~\ref{figure9}, except for $^{100-110}$Mo.}
\label{figure10}
\end{figure*}

\begin{figure*}
\centering
\includegraphics[scale=0.34]{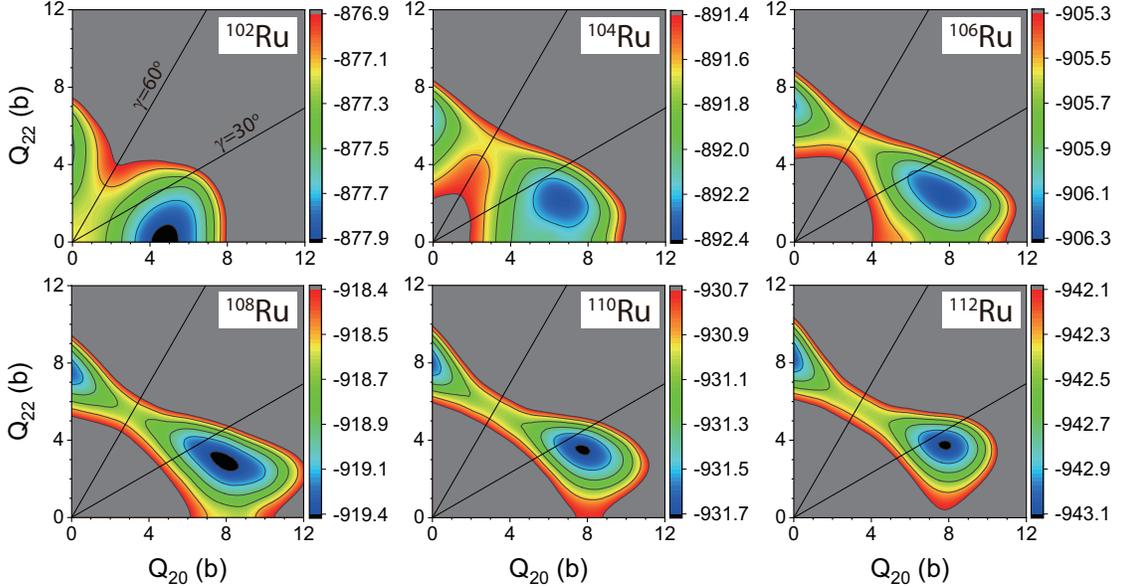}
	\caption{The same as Fig.~\ref{figure9}, except for $^{102-112}$Ru.}
\label{figure11}
\end{figure*}

Before showing the calculated IV $E1$ photoabsorption cross section of these
neutron-rich nuclei, it is necessary to have some idea about the
potential-energy surfaces of quadrupole deformations. Figures~\ref{figure9},
\ref{figure10}, and \ref{figure11} display the potential-energy surfaces for
even-even $^{98-108}$Zr, $^{100-110}$Mo, and $^{102-112}$Ru nuclei, calculated
with \UNEDFone~EDF. The constrained HFB calculations for these potential-energy
surfaces are performed with the \HFODD~code (version 2.68h~\cite{schunck17}).
For these HFB+LN calculations, there are 1140 ($N=17$) spherical HO bases
included; the original pairing strengths and energy cut-off on the
quasi-particle spectra are used~\cite{kort12}.

For the Zr isotopes, the ground states for $50\le N \le58$ are spherical due to
their closeness to the $Z=40$ subshell closure. For $^{100,102}$Zr ($N=60,62$),
the ground states show the coexistence of prolate and oblate minima, with the
prolate minimum being slightly lower energetically in $^{102}$Zr. For Zr
isotopes with $N\ge64$, the prolate minima move to a static triaxial
deformation, with the oblate minima staying slightly higher in energy.

The evolution of the minima of the neutron-rich Mo and Ru isotopes can be seen
in Figs.~\ref{figure10} and \ref{figure11}. In general, we observe a triaxial
minimum near prolately deformed region which is developed in the $N=60$
isotopes ($^{102}$Mo and $^{104}$Ru). This minimum increases $\gamma$ with
increasing neutron number. For all the isotopes with $58\le N\le 70$, the current
results predict a finite $\gamma$ deformation.

The shape evolution and shape coexistence near the ground states of the
neutron-rich zirconium isotopes are particularly interesting. With recent
advances in the rare isotope facilities, the experimental low-energy spectra
for the most neutron-rich isotopes in the Zr, Mo, and Ru
nuclei~\cite{naka17,wata11,doherty17} are becoming more and more available.  If
we examine the experimental and theoretical literatures, it is fair to say that
the current static calculations are in reasonable agreement with the
experimental data. The potential-energy surfaces obtained using UNEDF
parameters~\cite{zhang15} are somewhat more rigid in the $\gamma$ deformation
compared with theoretical calculations using other models and
parameters~\cite{skalski97, hilaire07, nomura16, togashi16, zhao17, miya18,
heyde19}. Thus, we continue our dynamic survey using \UNEDFone~EDF. The main
conclusions about the variations of GDR peaks due to quadrupole deformation
obtained here can be expanded to other parameters or even other mean-field
models.

\subsubsection{A case study: $^{100}$Mo}
\label{mo100}

In Sec.~\ref{spherical} we have shown that the current TDDFT + BCS calculations
give good description for the GDR peaks and the lower part of the strengths for
the spherical nuclei in this mass region. This section discusses the influence
of the deformation on the GDR peak of $^{100}$Mo.

\begin{figure}
\centering
\includegraphics[scale=0.5]{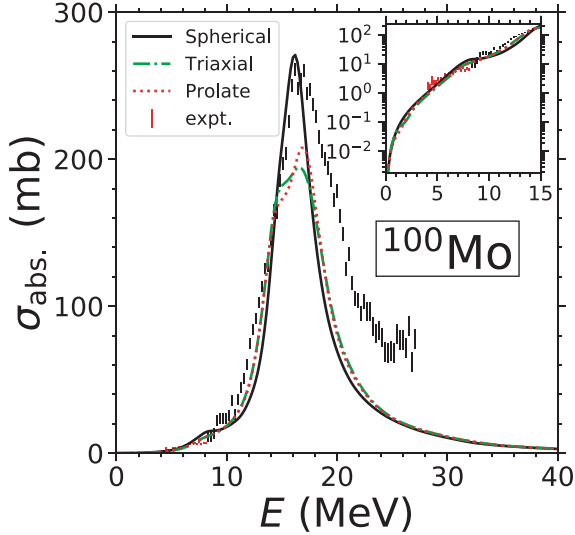}
	\caption{The cross sections of $^{100}$Mo calculated with
\UNEDFone~EDF, $\Gamma=2.0$\,MeV. The black solid line corresponds to the
spherical ground state as shown in Fig.~\ref{figure10}. The green dashed
line corresponds to the triaxial minimum in the energy surface of $^{100}$Mo,
with ($Q_{20}$, $Q_{22})$=(3.5,2.4)\,b. The above two curves are based on
results without constraints on the quadrupole moments. The red dotted line
corresponds to a prolate deformation which has been constrained to have
$Q_{20}=5.0$\,b. The experimental data are extracted from
Refs.~\cite{exfor,nndc}.}
\label{figure90}
\end{figure}

In Fig.~\ref{figure90} we show the calculated results of $^{100}$Mo using the
\UNEDFone~EDF and $\Gamma=2.0$\,MeV. The potential-energy surface of $^{100}$Mo
show significant softness around the spherical minimum, with a triaxial minimum
being very low in energy, see Fig.~\ref{figure10}. Hence, the cross-section
curves based on both minima are calculated and plotted in Fig.~\ref{figure90}.
In addition, a red dotted curve is included with the quadrupole moments being
artificially constrained to be $(Q_{20}, Q_{22})=(5.0,0.0)$\,b.

In Fig.~\ref{figure90}, we see that the calculated height of the GDR peak for
the spherical minimum reproduces that of the experimental data. The width has
been underestimated. The TRK sum rule value is 361.8 $e^2$\,fm$^2$\,MeV for
this nucleus. The EWSR from the TDDFT + BCS result is 415 $e^2$\,fm$^2$\,MeV
($\kappa=0.15$). The shape of the GDR peak based on the prolate deformation is
similar to that based on the triaxial deformation, except that the peak at the
higher energy is more pronounced. The relative heights of the peaks are related
to the deformation they are based on, as will be discussed in
Sec.~\ref{deformed_curve}. The cross sections calculated based on the three
minima well reproduce the low-energy ($E<10$\,MeV) part of the experimental
data. Comparing our calculated GDR peaks based on spherical, prolate, and
triaxial minima with that of the experimental data, it seems that the data
support an explanation that the IVD vibration is based on a spherical minimum.

\subsubsection{Dynamical results for Zr isotopes: shape coexistence}
\label{deformed_curve}

\begin{table}[htb]
	\caption{The ground-state (and the coexisting minima for Zr isotopes)
properties, the pairing energies, the quadrupole moment, and the
triaxial parameter ($\gamma$) of $^{100-108}$Zr, $^{102-110}$Mo, and $^{104-112}$Ru
calculated with the HF+BCS code with the finite-difference method.
The $Q_{22}$ value is connected with the $Q_{20}$ and $\gamma$ values through
$Q_{22}=Q_{20}\tan\gamma$.
The pairing strengths for neutrons and protons are $-$382 and $-$440 MeV\,fm$^3$,
respectively.}
\label{table4}
\begin{ruledtabular}
\begin{tabular}{ccccccc}
Nuclei       & $E_{\rm pair}^{\rm n}$ & $E_{\rm pair}^{\rm p}$       & $Q_{20}$  &  $\gamma$        & $m_1^{\rm TDDFT}$ & $m_1^{\rm g.s.}$  \\
\cline{2-7}
             &             \multicolumn{2}{c}{(MeV)}                 &    (b)    &  (deg)     &        \multicolumn{2}{c}{($e^2$\,fm$^2$\,MeV)}  \\
	\hline
\multirow{2}{*}{$^{100}$Zr} & $-$4.842 &  $-$3.778 & 2.286 & 60 & 410.3 &  420.7  \\
                            & $-$2.676 &  $-$3.488 & 9.894 & 0  & 410.4 &  421.0  \\
\multirow{2}{*}{$^{102}$Zr} & $-$7.846 &  $-$3.541 & 2.591 & 60 & 415.1 &  426.0  \\
                            & $-$3.775 &  $-$3.435 & 10.282 & 0 & 416.3 &  426.4  \\
\multirow{2}{*}{$^{104}$Zr} & $-$2.223 &  $-$2.711 & 9.555 & 16 & 411.9 &  431.6  \\
                            & $-$3.854 &  $-$3.173 & 2.502 & 60 & 423.8 &  431.1  \\
\multirow{2}{*}{$^{106}$Zr} & $-$1.717 &  $-$2.457 & 9.352 & 19 & 439.2 &  436.6  \\
                            & $-$2.543 &  $-$3.132 & 2.943 & 60 & 426.8 &  436.1  \\
\multirow{2}{*}{$^{108}$Zr} & $-$0.670 &  $-$2.450 & 9.102 & 19 & 444.9 &  441.3  \\
                            & $-$1.165 &  $-$3.024 & 3.161 & 60 & 429.8 &  441.0  \\
\hline
$^{102}$Mo    & $-$4.686 & $-$3.362 & 8.454 & 15 & 422.9 & 433.6   \\
$^{104}$Mo    & $-$4.178 & $-$2.984 & 9.100 & 16 & 428.6 & 439.5   \\
$^{106}$Mo    & $-$3.865 & $-$2.624 & 10.010 & 16 & 433.4 & 445.0  \\
$^{108}$Mo    & $-$4.240 & $-$2.369 & 9.759 & 18 & 437.9 & 450.2   \\
$^{110}$Mo    & $-$4.007 & $-$2.322 & 9.200 & 23 & 453.0 & 455.3   \\
\hline
$^{104}$Ru    & $-$5.192 & $-$2.894 & 7.871 & 16 & 433.9 & 445.8   \\
$^{106}$Ru    & $-$5.473 & $-$2.777 & 8.110 & 16 & 438.2 & 452.0   \\
$^{108}$Ru    & $-$5.953 & $-$2.271 & 7.957 & 18 & 447.7 & 457.9   \\
$^{110}$Ru    & $-$5.363 & $-$2.116 & 8.442 & 19 & 450.7 & 463.6   \\
$^{112}$Ru    & $-$5.113 & $-$1.716 & 8.160 & 24 & 455.7 & 468.9   \\
\end{tabular}
\end{ruledtabular}
\end{table}

\begin{figure*}
\centering
\includegraphics[scale=0.8]{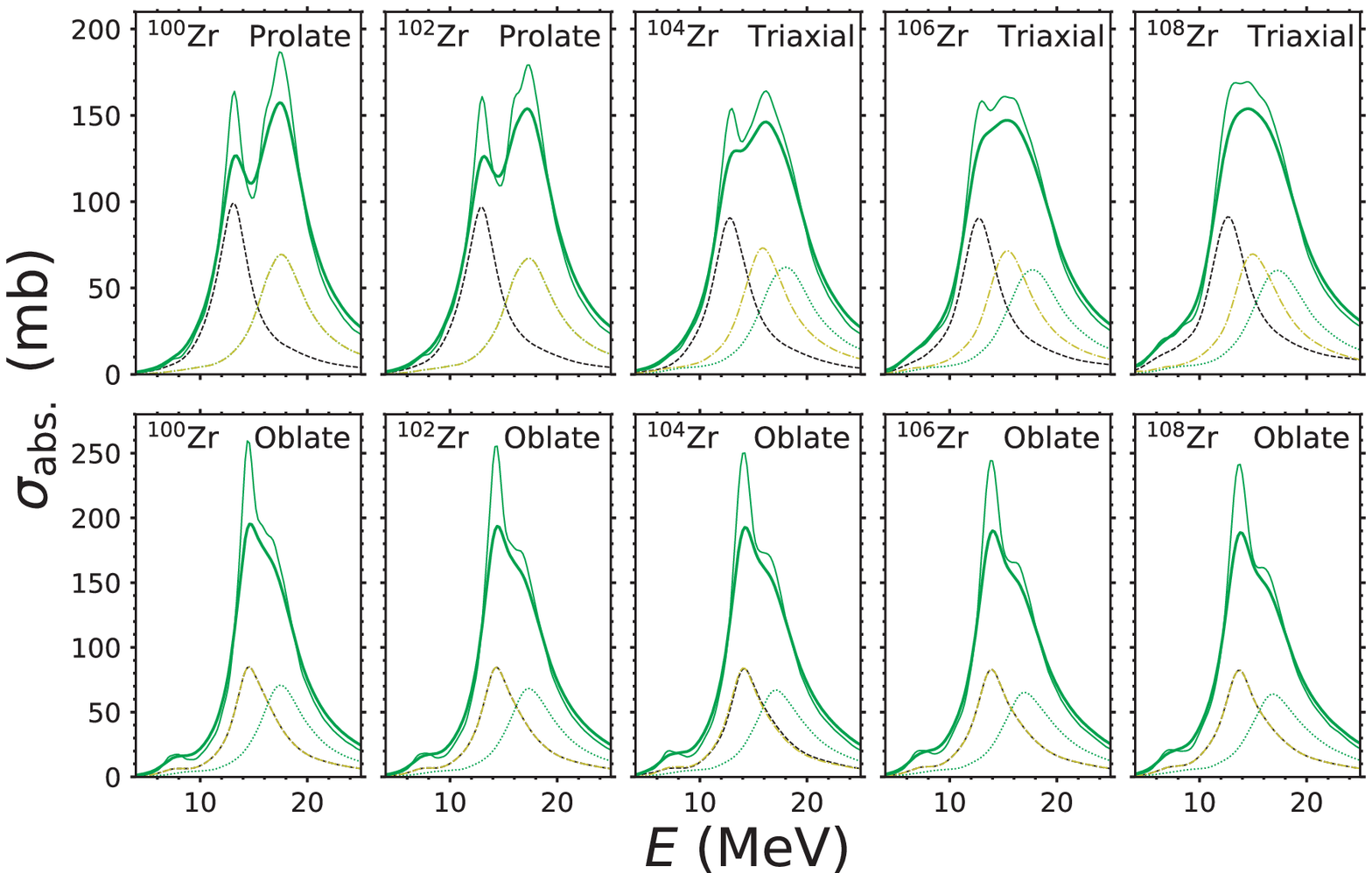}
	\caption{Photoabsorption cross sections calculated for $^{100-108}$Zr with
\UNEDFone~EDF with smoothing parameter $\Gamma= 2.0$\,MeV (thick line) and 1.0 MeV
(thin line). The plots in upper row correspond to the results based on the
prolate minima for $^{100,102}$Zr, and the triaxial minima for $^{104-108}$Zr;
the plots in the lower row correspond to those based on the oblate minima, see
Fig.~\ref{figure9}. The thinner green lines indicate results with
$\Gamma=1.0$\,MeV.}
\label{figure12}
\end{figure*}

The systematic calculations of TDDFT+BCS for Zr, Mo, and Ru isotopes are
displayed in Figs.~\ref{figure12} and \ref{figure13}.  The relevant information
on the static results before the time propagation is listed in
Table~\ref{table4}. These HF+BCS calculations using the finite-difference
method are without deformation constraints.

For the triaxially deformed minima in $^{104-108}$Zr, $^{102-110}$Mo, and
$^{104-112}$Ru, the HF+BCS calculations give smaller $\gamma$ compared to that
of the HFB+LN results using \HFODD~(Figs.~\ref{figure9}, \ref{figure10}, and
\ref{figure11}). Specifically, for the softest nucleus, the triaxially deformed
$^{110}$Mo, the HFB+LN calculation using \HFODD~gives quadrupole moments
$(Q^{\rm HFB+LN}_{20}, Q^{\rm HFB+LN}_{22})\approx(8.5,3.5)$\,b (see
Fig.~\ref{figure10}). Without pairing, the \HFODD~calculation gives
$(Q^\textrm{HF}_{20}, Q^\textrm{HF}_{22})\approx(9.4,4.0)$\,b. Before
performing dynamic calculation, the static HF+BCS calculation in the Cartesian
coordinate space gives $(Q_{20}, Q_{22})\approx(9.2,3.9)$\,b.

\begin{figure}
\centering
\includegraphics[scale=0.4]{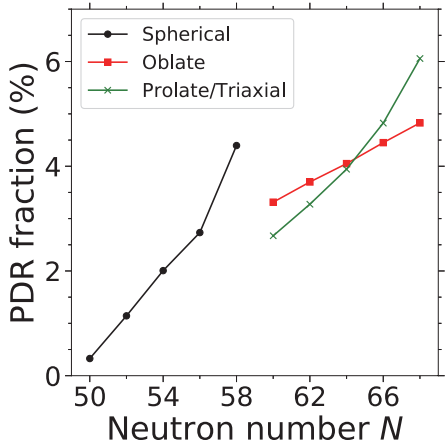}
	\caption{The fraction of the strengths for IVD resonances below
$E_c=10$\,MeV [Eq.~(\ref{fraction})] for zirconium isotopes with spherical,
prolate(for $^{100,102}$Zr)/triaxial(for $^{104,106,108}$Zr), and oblate
deformations. The smoothing parameter $\Gamma=0.5$\,MeV is used in the calculation. 
The integrations for the total strengths are taken from 0 to 80 MeV.}
\label{figure14}
\end{figure}

For $^{100,102}$Zr shown in Fig.~\ref{figure12} the GDR peaks split into two,
with the sub-peak at a slightly lower energy ($E\approx13$\,MeV) and the height of
the peak is lower than that at the higher energy ($E\approx18$\,MeV). For
$^{104}$Zr, the IVD vibrations are based on a weakly triaxial minimum, see
Fig.~\ref{figure12}.  The peaks at the higher excitation energies become
broader. For $^{106,108}$Zr, the two peaks merge to form one broader peak.

The features of the GDR peaks can be understood qualitatively. For a spherical
nucleus, the GDR peaks corresponding to the three vibrational modes are
identical due to the spherical symmetry. When the nucleus acquires an
axially-symmetric deformation, the GDR peaks split into two groups: (1) a mode
corresponding to a vibration along the symmetry axis ($K=0$ mode); and (2) two
modes corresponding to the vibrations along the axes perpendicular to the
symmetry axis ($K=\pm1$ modes). For a prolate shape, intuitively, because of
the larger material extension, the potential is enlongated along the symmetry
axis.  Hence, the energy cost is lower for the $K=0$ mode, compared to the
$K=\pm1$ modes. While the peak for the $K=0$ mode shifts to a lower energy, the
contribution of this mode to the total strength becomes larger than those of
the $K=\pm1$ modes. Similar effects can be found from light to heavy spherical
nuclei, where the total GDR peak shifts to a lower energy and the strength
becomes larger and/or broader.

For these neutron-rich Zr isotopes, the oblate minima appear at relatively low
energies. As discussed above, the peaks corresponding to the two longer axes
($K=\pm 1$) appear at lower energies, and the strengths are larger compared to
that from the shorter axis ($K=0$). This results in the peaks of the total
cross section at the lower energies ($E\approx14$\,MeV) considerably higher than
those at higher energies ($E\approx18$\,MeV) and higher than those of the prolate
deformation. Because of the large smoothing parameter ($\Gamma=2.0$\,MeV), the
second peaks appear to be shoulders of the first higher peaks for these nuclei.

Figure~\ref{figure14} plots the PDR fraction $f_{\rm PDR}$ below $E_c=10$\,MeV
for Zr isotopes based on different deformations. With the same deformation, the
$f_{\rm PDR}$ values increase with neutron excess. From spherical to deformed
nuclei, we see a small decrease of the $f_{\rm PDR}$ value at $N=60$, which
agrees with the previous studies~\cite{inak11,ebata14}. For the case of the
transition from a spherical to a prolate deformation, this is a net result of
(1) the decrease of energy of the $K=0$ mode, and the increase of the energy of
the $K=\pm1$ modes, as well as (2) an enhanced contribution in the total
strength from the $K=0$ mode, as pointed out in Ref.~\cite{arteaga09}. For the
transition from a spherical to an oblate deformation, similar effects are also
responsible for the decrease of the PDR fraction. For the oblate deformation,
there is a plateau structure below $E=10$\,MeV, which is the main contribution
to the $f_{\rm PDR}$ value. The slope of $f_{\rm PDR}$ curve are smaller for
the oblate deformation compared to that of the prolate deformation. This is
because the oblate deformations are relatively constant, whereas the prolate
minima become weak triaxial with increasing neutron number.

\subsubsection{Dynamical results for Mo and Ru isotopes: triaxial deformation}
\label{mo_ru}

\begin{figure*}
\centering
\includegraphics[scale=0.8]{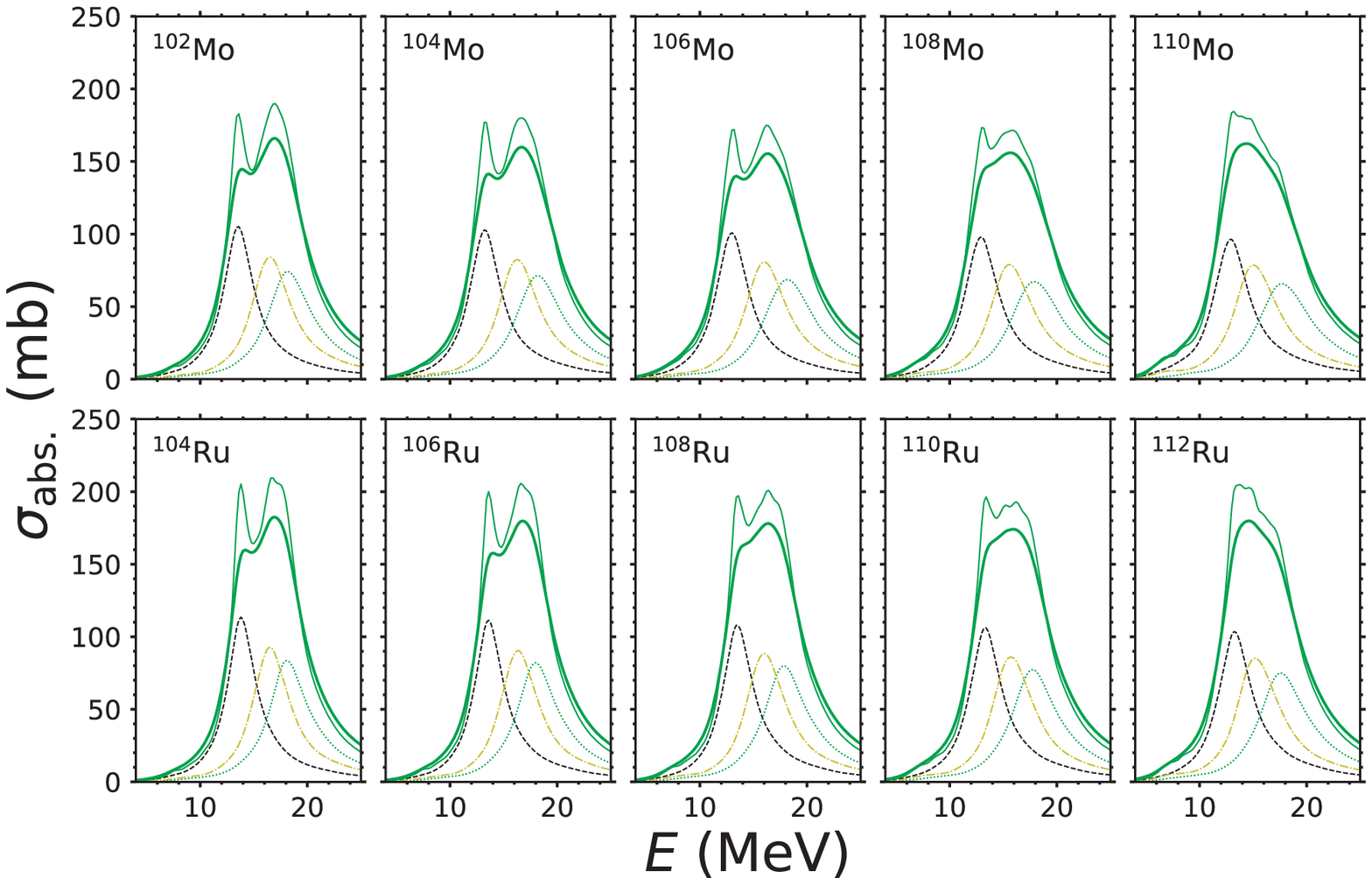}
	\caption{The same as Fig.~\ref{figure12}, except for $^{102-110}$Mo and
$^{104-112}$Ru.}
\label{figure13}
\end{figure*}

Figure~\ref{figure13} plots the IVD cross sections calculated for the Mo and Ru
isotopes with $60\le N\le 68$. In the current work, these nuclei are calculated
to be triaxial. The properties concerning the ground states are listed in
Table~\ref{table4}. For $^{102-106}$Mo, the GDR peaks appear to include two
sub-peaks instead of three, due to the weak $\gamma$ deformation and large
$\Gamma$. For $^{110}$Mo, the three peaks due to the vibrations in the three
Cartesian directions merge into one broad peak. It is interesting to note that
from $^{100}$Mo to $^{102}$Mo one sees a transition in deformation from soft
spherical to a soft triaxial shape, see Fig.~\ref{figure10}. The total cross
section for $^{100}$Mo has been discussed in Sec.~\ref{mo100}.

For deformed nuclei, the experimental data seem to indicate a smaller smoothing
parameter compared to that of the spherical ones~\cite{oishi16}. Thus, in
Figs.~\ref{figure12} and \ref{figure13}, we plot the same total cross sections
with a smaller $\Gamma=1.0$\,MeV. We can see that, as expected, the heights of
the GDR peaks are larger compared to those with $\Gamma=2.0$\,MeV. With better
resolution, we can see more detailed structures due to the different subpeaks.

\begin{figure}
\centering
\includegraphics[scale=0.4]{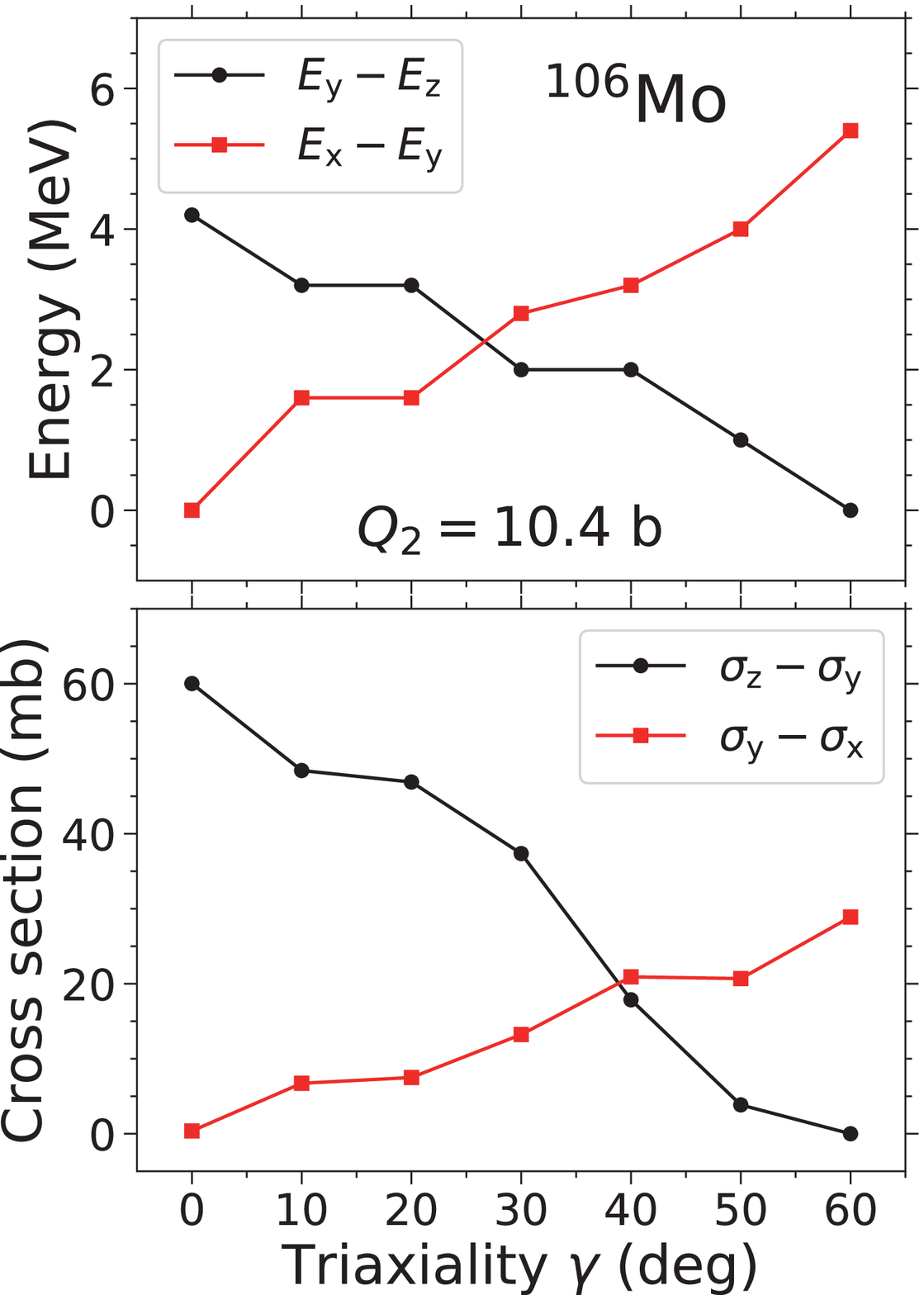}
	\caption{The upper pannel are the energy differences of peaks corresponding
to the vibrations along the $y-$ and $z-$axis ($E_y-E_z$), as
well as along the $x-$ and $y-$axis for $\gamma=0^{\circ}-60^{\circ}$ of
$^{106}$Mo. The lower pannel are the calculated relative cross sections corresponding
to the energies in the upper pannel, that is, 
$\sigma_{\rm x,y,z}\equiv\sigma_{\rm abs.}(E_{\rm x,y,z})$.}
\label{figure15}
\end{figure}

Figure~\ref{figure15}(a) plots the energy differences corresponding to the
three GDR peaks due to the $K$ modes along the three axes. The
$Q_2\equiv\sqrt{Q_{20}^2+Q_{22}^2}$ is constrained to be 10.4\,b while $\gamma$
deformation is constrained to vary from 0$^{\circ}$ to 60$^{\circ}$. It can be
seen that the energy difference between $E_y$ and $E_z$, which corresponds to
the vibration along the long and medium axes, respectively, decreases
monotonically as $\gamma$ increases from a prolate to an oblate deformation.
Whereas the energy difference between $E_x$ and $E_y$ increases with $\gamma$
deformation. For axial deformations, the energy split of the two peaks are
larger for the oblate deformation than that of the prolate deformation.

Figure~\ref{figure15}(b) shows the relative heights of the GDR cross sections
corresponding to the vibration modes along the three axes as a function of the
triaxial deformation. We see that the height corresponding to the vibration of
the long axis ($z$-axis) is always larger than that of the medium axis
($y$-axis). Similarly, $\sigma_{\rm abs.}(E_{\rm y})$ is always larger than
$\sigma_{\rm abs.}(E_{\rm x})$. This can be qualitatively explained: the
further the material extends the lower the peak energy becomes; the lower the
energy becomes, the larger the height becomes. With inreasing $\gamma$,
$\sigma_{\rm abs.}(E_{\rm z})-\sigma_{\rm abs.}(E_{\rm y})$ decreases
continuously until $\sigma_{\rm abs.}(E_{\rm z})=\sigma_{\rm abs.}(E_{\rm y})$
for an oblate deformation.

We end this section by emphasizing the interests associated with the current
study. The experimental signature of the triaxial deformation in the nuclear
ground state is not well established, which is mainly due to the theoretical
challenges in uniquely connecting the spectroscopic observables with the
triaxial degree of freedom. The investigations presented in this section may
provide prospects to establish a firm connection between the photoabsorption
cross section data and the ground-state triaxial deformation. Indeed, for the
heavier Mo and Ru isotopes, if future experiments allow for resolving the
general shapes (peak heights and splittings) of the individual peaks originated
from different $K$ modes, then it is possible to determine the triaxiality
parameter of the ground state by comparing the experimental plot similar to
Fig.~\ref{figure15}.

\section{Summary} 
\label{summary}

Based on a previous computer code developed for the nuclear density-functional
theory (DFT), we present a further development, enabling the time-dependent DFT
(TDDFT) calculations with BCS pairing. We benchmark the code by comparing its
calculated response functions of the dipole moment of $^{16}$O with that of an
existing 3D TDDFT code, Sky3D. Although the response functions for $^{16}$O are
sensitive to a few subtle factors (time-odd mean fields, treatment of boundary
conditions, etc.), a remarkable agreement has been found between the two codes,
as long as those factors are carefully considered.

To apply the TDDFT + BCS in its linearized limit and describe the isovector
(IV) electric dipole ($E1$) observables, we carry out finite-amplitude method
for quasiparticle random-phase approximation (FAM-QRPA) for a few light
spherical ($^{16}$O, $^{40}$Ca) and axially deformed ($^{24,34}$Mg) nuclei, and
compare the calculated IV $E1$ properties with those resulted from the TDDFT +
BCS calculations. The comparisons are acceptable up to the first peak at
$E\approx20$\,MeV. Beyond that, the FAM-QRPA calculations based on the
harmonic-oscillator basis give more fragmented peaks compared to that of the
TDDFT + BCS calculations employing the absorbing boundary condition.

Using the \UNEDFone~energy density functional (EDF), the current TDDFT + BCS
calculations provide reasonable descriptions for both the giant dipole
resonance and the low-energy part of the IV $E1$ photoabsorption cross section
for spherical Zr and Mo isotopes, where experimental data exist. 

For heavier Zr isotopes, the calculated potential-energy surfaces show
coexisting minima. The predicted $E1$ photoabsorption cross sections reflect
typical features depending on the local minima that they are based upon. 

For heavier Mo and Ru isotopes, the ground states are predicted to be triaxial.
The predicted cross sections show features that distinguish them from the
spherical ones. For Mo isotopes considered here, the predicted onset of the
triaxial deformation which occurs in $^{102}$Mo ($N=60$), is only two neutrons
larger than the isotope, $^{100}$Mo, in which experimental data exist. The
systematic measurements of the photonuclear experiments on these Mo isotopes
are desired for further analysis of the ground-state triaxial deformation.

\begin{acknowledgments}

Useful discussions with T. Nakatsukasa, W. Nazarewicz and P. Stevenson are gratefully acknowledged.
The current work is supported by National Natural Science Foundation of China (Grant No. 11705038, No. 12075068),
JSPS KAKENHI Grant No. 16K17680, No. 20K03964, 
the JSPS-NSFC Bilateral Program for the Joint Research Project on ``Nuclear mass and life for unravelling mysteries of r-process'',
and the Deutsche Forschungsgemeinschaft (DFG, German Research Foundation) - Projektnummer 279384907 - SFB 1245.
YS thanks the HPC Studio at Physics Department of Harbin Institute of 
Technology for computing resources allocated through INSPUR-HPC@PHY.HIT.
A part of the numerical calculations were performed at the Oakforest-PACS 
Systems through the Multidisciplinary Cooperative Research Program of the 
Center for Computational Sciences, University of Tsukuba.

\end{acknowledgments}

%

\end{document}